\documentclass[10pt,twocolumn,letterpaper]{article}

\usepackage{array}
\usepackage{overpic}
\usepackage{float}
\usepackage{rotating}
\usepackage[caption=false]{subfig}
\usepackage{multirow}
\usepackage{amssymb}
\usepackage{algorithm}
\usepackage{algorithmic}
\usepackage[cmyk]{xcolor} 
\usepackage[pagenumbers]{cvpr} 

\definecolor{cvprblue}{rgb}{0.21,0.49,0.74}
\usepackage[pagebackref,breaklinks,colorlinks,citecolor=cvprblue]{hyperref}

\def\paperID{1263} 
\def\confName{CVPR}
\def\confYear{2024}

\title{Computational Spectral Imaging with Unified Encoding Model: \\ A Comparative Study and Beyond}

\author{
Xinyuan Liu \qquad Lizhi Wang\thanks{Corresponding author: Lizhi Wang.} \qquad Lingen Li\\ 
Beijing Institute of Technology\\
\and
Chang Chen \qquad Xue Hu \qquad Fenglong Song \qquad Youliang Yan\\
Huawei Noah's Ark Lab\\
}

\begin{document}
\maketitle
\begin{abstract}
Computational spectral imaging is drawing increasing attention owing to the snapshot advantage, and amplitude, phase, and wavelength encoding systems are three types of representative implementations. Fairly comparing and understanding the performance of these systems is essential, but challenging due to the heterogeneity in encoding design.
To overcome this limitation, we propose the unified encoding model (UEM) that covers all physical systems using the three encoding types. Specifically, the UEM comprises physical amplitude, physical phase, and physical wavelength encoding models that can be combined with a digital decoding model in a joint encoder-decoder optimization framework to compare the three systems under a unified experimental setup fairly.
Furthermore, we extend the UEMs to ideal versions, namely, ideal amplitude, ideal phase, and ideal wavelength encoding models, which are free from physical constraints, to explore the full potential of the three types of computational spectral imaging systems.
Finally, we conduct a holistic comparison of the three types of computational spectral imaging systems and provide valuable insights for designing and exploiting these systems in the future.
\end{abstract}

\section{Introduction}
\label{sec:intro}
Spectral imaging is a valuable technique that captures both spatial and spectral information of a scene along a specific range of wavelengths in the electromagnetic spectrum.
The analysis of the information facilitates material discrimination and object classification.
Consequently, spectral imaging has found applications in various fields, such as biology~\cite{li2013review}, medicine~\cite{zeng2022research}, and agriculture~\cite{mancini2019challenges}. 
Conventional spectral imaging systems are bulky, require time-consuming scanning to acquire spatial and spectral information, and are difficult to use in dynamic scenes.
In contrast, computational spectral imaging systems capture spectral scenes into a single snapshot measurement using well-designed optical encoders and then reconstruct them using corresponding digital decoders. 
Therefore, computational spectral imaging is the most promising solution to capture spectral information of dynamic scenes without scanning efficiently.

Currently, physically implementable computational spectral imaging systems can be divided into three representative categories based on the encoding method.
Amplitude encoding systems based on compressive sensing~\cite{gehm2007single, august2013compressive, llull2013coded}, phase encoding systems that rely on diffractive optics~\cite{golub2016compressed, 2019Compact, hauser2020dd}, and wavelength encoding systems that employ filters~\cite{nguyen2014training, robles2015single}. 
Research on the three kinds of computational spectral imaging systems has been limited to specific categories, resulting in the absence of a comprehensive understanding of the technology.
Moreover, due to inherent differences among these systems, artificial optics specifically designed for one cannot be transferred to another, thus presenting challenges for fair comparisons.
Therefore, conducting comprehensive, impartial, and inclusive comparative investigations within a unified context is urgently needed to gain a broader perspective and thorough understanding of these systems.

In this paper, we propose the Unified Encoding Model (UEM) to comprehensively compare the performance of the three types of spectral imaging systems under a unified framework.
Representative spectral imaging systems can be modeled using physical coding techniques such as amplitude (\mbox{AEM-P}), phase (\mbox{PEM-P}), or wavelength (\mbox{WEM-P}).
We choose three typical physical systems: CASSI as \mbox{AEM-P}, Rotationally Symmetric DOE as \mbox{PEM-P}, and Selected Filter as \mbox{WEM-P}.

Moreover, to fully explore the potential performance of the three types of computational spectral imaging systems in our comparative study, we propose ideal models with greater degrees of freedom that eliminate the constraints of physical models.
These ideal models, including the ideal amplitude encoding model (\mbox{AEM-I}), ideal phase encoding model (\mbox{PEM-I}), and ideal wavelength encoding model (\mbox{WEM-I}), offer an opportunity to push the performance limits of each type of spectral imaging system.
The \mbox{AEM-I} model enables the free modulation of amplitude to any possible value (i.e., 32-bit floating point number) in both spatial and spectral dimensions, the \mbox{PEM-I} model enables any form of the point spread function (PSF), and the \mbox{WEM-I} model allows any shape of the spectral response curve.
The encoding properties obtained through the high-degree-of-freedom optimization process provide valuable guidance for computational spectral imaging.

Furthermore, We construct an end-to-end computational spectral imaging framework consisting of the UEMs and three types of decoding models. To ensure a fair comparison between the three types of computational spectral systems, we employ a joint encoder-decoder optimization approach~\cite{sitzmann2018end} to optimize the three casted UEMs and corresponding decoding models for the same spectral imaging objective.

Additionally, to explore the performance of computational spectral imaging on high-level tasks, we conduct spectral image segmentation experiments and observe no positive correlation between segmentation performance and imaging quality.
By characterizing each system in the experiment, we gain a more comprehensive understanding of the three computational spectral imaging systems and guide the development of computational spectral imaging systems.

In summary, our contributions are as follows:

\begin{itemize}[leftmargin=7mm, itemindent=0mm]
	\item We propose the UEM for the three physically implementable computational spectral imaging systems, enabling a fair comparison under the unified framework.
	
	\item We extend the UEM to the ideal version with increased degrees of freedom, eliminating the physical constraints of the three models. The ideal UEM allows for a thorough exploration of the potential performance of the three types of computational spectral imaging systems.
	
	\item We construct a computational spectral imaging framework with the UEMs and decoding models, using a joint encoder-decoder optimization approach to compare three types of computational spectral systems.
        
        \item We investigate the impact of low-level imaging performance on downstream high-level segmentation tasks.
	
\end{itemize}

\section{Related Work}

\noindent {\bf{Amplitude Encoding System.}} 
The amplitude encoding system is based on the theory of compressive sensing~\cite{donoho2006compressed,wakin2006architecture,candes2008introduction}, which combines optics, mathematics, and optimization theory. This system employs an encoding mask to block or filter the input light and then passes through the dispersion element to acquire an image with compressed information~\cite{duarte2008single,duarte2013spectral,arce2013compressive,martin2014hyca,cao2016computational,oiknine2018compressive}. The image generation is expressed as a point multiplication process between a mask that is consistent with the image size and a monochromatic object image for each wavelength. Finally, a specific reconstruction algorithm decodes and obtains the underlying spectral image.

HyperReconNet~\cite{wang2018hyperreconnet} jointly learns the coded mask and the corresponding CNN for reconstruction. The BinaryConnect~\cite{courbariaux2015binaryconnect} adds the encoded mask to the network as a layer, using a floating-point mapping to the binary-encoded mask entity.
E. Salazar et al.~\cite{salazar2020coded} proposed an optimal coded mask algorithm for the Spatial Spectral Compressive Spectral Imager (SSCSI) based on a discrete measurement model of his design. 
D2UF~\cite{jacome2022d} encoded the mask and multispectral color filter array of the coded aperture snapshot spectral imager (CASSI) and performed spectral reconstruction using an unrolling-based network.

\begin{figure*}[!t]
	\centering
	\includegraphics[width=6.5in]{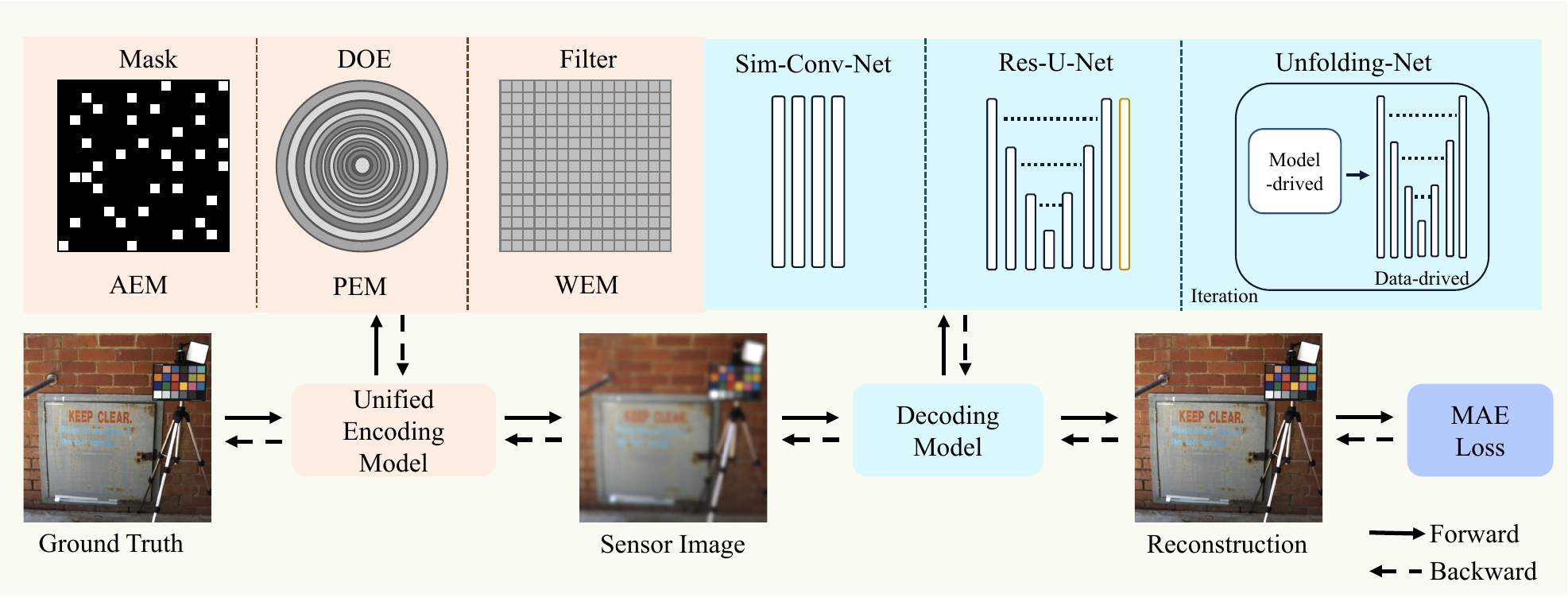}
	\caption{Joint encoder-decoder optimization framework optimizes the UEM and the decoding model toward the same objective.}
	\label{fig_joint_optimization_framework}
\end{figure*}

\noindent {\bf{Phase Encoding System.}} 
The phase encoding system is based on diffractive imaging~\cite{hauser2020dd}, employing diffractive optical elements(DOE)~\cite{taghizadeh1997design} or metasurfaces~\cite{neshev2023enabling}.
These elements are specifically designed to control the phase modulation in terms of the PSF to act as encoders that collect desired spectral information.

Several research efforts have been made to improve the phase encoding system. Peng et al.~\cite{peng2016diffractive} proposed a shape-invariant PSF design approach for high-quality color imaging. In contrast, Jeon et al.~\cite{jeon2019compact} designed a spectrally varying PSF that rotates regularly with wavelength and encodes spectral information. Xiong et al.~\cite{dun2020learned} decomposed the PSF formula into Bessel functions, simplified these functions according to rotational symmetry, and reconstructed the network using a W-Net composed of Res-U-Net units. Arguello et al.~\cite{arguello2021shift} introduced a color-coded aperture and designed the spatially variable PSF.

To address the quantization issue of DOE height maps, Li et al.~\cite{li2022quantization} integrated the quantization operation into the DOE optimization process, using an adaptive mechanism to optimize the physical height of each volume layer. Tseng et al.~\cite{tseng2021neural}, and Makarenko et al.~\cite{makarenko2022real} used metasurface instead of DOE as the encoding element, achieving excellent results.

\noindent {\bf{Wavelength Encoding System.}} 
The wavelength encoding system uses RGB narrowband filters or other band filters directly to encode the scene in the wavelength dimension and perform spectral reconstruction of the captured image.~\cite{glassner1989derive,sun1999deriving,smits1999rgb} Wavelength encoding systems produce better spatial quality in reconstructed results due to the absence of modulation on spatial dimensions.

The prevalent wavelength encoding system directly regards the imaging pipeline of an arbitrary or well-selected RGB camera as the encoding process, constructing reconstruction algorithms from RGB images.
Xiong et al.~\cite{xiong2017hscnn} proposed HSCNN, which applied CNN to upsampled RGB and amplitude-encoded measurements for spectral reconstruction.
Galliani et al. ~\cite{galliani2017learned} proposed a learned spectral super-resolution using CNN for end-to-end RGB to spectral image reconstruction.
The current popular attention mechanism~\cite{vaswani2017attention} has also been applied in spectral imaging. Cai et al. ~\cite{cai2022mst++} proposed Transformer-based MST++, which uses cells based on HSI spatial sparsity and spectral self-similarity properties as the base module.

Admittedly, existing RGB narrowband filters are designed to satisfy human vision and are intuitively sub-optimal for spectral imaging. Therefore, some researchers chose filters to achieve a suitable spectral response for spectral imaging~\cite{fu2020joint}, and some optimized the response to obtain better spectral imaging performance.
Nie et al.~\cite{nie2018deeply} encoded the RGB filter and combined this function with the reconstruction network for end-to-end optimization on the spectral reconstruction task.

\noindent {\bf{Joint Encoder-decoder Optimization.}}
With the advancements in computational photography, joint encoder-decoder optimization methods have become the mainstream technique to enhance imaging performance.
Instead of designing the encoder and the decoder separately like traditional approaches, the joint encoder-decoder optimization algorithm optimizes both the encoder and decoder simultaneously for a given task.
The superior performance of the joint encoder-decoder optimization method has been demonstrated in various computational imaging tasks such as HDR imaging~\cite{metzler2020deep, sun2020learning}, depth estimation~\cite{ikoma2021depth, baek2021single, liu2022investigating} and computational spectral imaging~\cite{dun2020learned, li2022quantization}. 
However, joint encoder-decoder optimization in computational spectral imaging is still limited to specific categories, lacking comprehensive understanding.

\section{Unified Encoding Model}

\subsection{Overview}

We propose the UEM that covers all the possible types of optical modulations for computational spectral imaging, which can be cast into AEM, PEM, and WEM to represent amplitude, phase, and wavelength encoding systems, respectively. 
Specifically, for the amplitude modulation, the scene light is element-wise multiple with the matrix $A(x,y,\lambda)$ of the encoding mask; 
For phase modulation, the light is convolved with the PSF $P(x,y,\lambda)$ of the DOE; 
For wavelength modulation, the light is multiplied by the spectral response $W(\lambda)$ of the filter, integrated along the wavelength dimension $\lambda$, and converted into a measurement $I_{rgb}$ on the RGB sensor. 
We give the overall formula for the UEM as follows, with some terms becoming unit terms under specific settings:
\begin{equation}
	I_{rgb} = \int W(\lambda) \cdot P(x,y,\lambda) \ast (A(x,y,\lambda)\cdot I(x,y,\lambda) ) d\lambda,
	\label{eq:uem}
\end{equation}	
where $I(x,y,\lambda)$ denotes the natural spectral scene, and $I_{rgb}$ is the RGB encoded image captured by the sensor.

In the following, we introduce two versions of UEMs: the physical and the ideal versions.
The physical UEM is modeled with physical constraints in the encoding process, corresponding to a series of physically implementable encoding systems for spectral imaging.
Based on the physical UEM, we can put the three types of physically implementable spectral imaging systems into one unified setting for comparative study.
Beyond, the ideal UEMs are extended from the physical UEMs by eliminating the physical constraints of the encoding process, which means the ideal UEMs have higher degrees of encoding freedom. 
We employ the ideal UEMs to investigate the maximum potential of each type of spectral imaging system. 
For convenience, we use the suffix `-P' to denote the physical UEMs and the suffix `-I' to indicate ideal UEMs.

\begin{table}[!t]
	\centering
	\caption{Physical systems corresponding to each physical UEM.}
	\label{table_model}
	\resizebox{\linewidth}{!}
	{\begin{tabular}{cc}
			\hline
			\textbf{Encoding Model} & \textbf{Physical Systems} \\ \hline
			\textbf{AEM-P} & CASSI~\cite{wagadarikar2008single,lin2014dual, correa2014compressive, arguello2014colored}, SSCSI~\cite{salazar2020coded} \\ 
			\textbf{PEM-P} & Rotationally Symmetric DOE~\cite{dun2020learned}, Free DOE~\cite{baek2021single} \\ 
			\textbf{WEM-P} & Selected Filter~\cite{2020Joint}, Learned Filter~\cite{nie2018deeply} \\ \hline
	\end{tabular}}
\end{table}

\subsection{Physical Model}
We first introduce the physical UEMs and cast them to \mbox{AEM-P}, \mbox{PEM-P}, and \mbox{WEM-P} for physically implementable amplitude, phase, and wavelength encoding systems, respectively.
The representative physical systems that correspond to each physical UEM are presented in Table~\ref{table_model}.

\noindent {\bf{Physical Amplitude Encoding Model.}} 
The \mbox{AEM-P} only employs the amplitude modulation part of the UEM, with physical constraints on the modulation that should be implementable using a mask.

CASSI~\cite{wagadarikar2008single, august2013compressive, llull2013coded,correa2015snapshot,lin2014spatial,rueda2014compressive,sun2021unsupervised} is a widely used physical amplitude encoding system. 
The system uses a binary encoding mask to modulate the spatial amplitude of the incident light, followed by a dispersion element that introduces distinct offsets for varying wavelengths of light, ultimately compressing the spectral information along the spatial dimension.
Specifically, the $A(x,y,\lambda)$  of \mbox{AEM-P} encodes the unit shift of the mask upwards layer by layer in the spectral dimension.
\begin{equation}
	A(x,y,\lambda) = Mask_{P}(x,y,\theta_{mask_p}) \cdot Dispersive(\lambda),
\end{equation}	
where, the $Mask_{P}$ denotes the spatial dimensional encoding mask, the $\theta_{mask_p}$ is a learnable parameters, and the $Dispersive(\lambda)$ represents the dispersion operation.

The imaging process of \mbox{AEM-P} can be written based on UEM as follows:
\begin{equation}
	I_{rgb} = \int W(\lambda) \cdot (A(x,y,\lambda)\cdot I(x,y,\lambda) ) d\lambda.
	\label{aemp_imaging}
\end{equation}

\noindent {\bf{Physical Phase Encoding Model.}} 
The \mbox{PEM-P} controls the height map or metasurface structure of the DOE to produce different PSFs, which describe the image-blurring effect by point light source. The resulting phase delay is used as phase modulation.
Following the standard settings, we model \mbox{PEM-P} based on a rotationally symmetric DOE, which uses a one-dimensional parameter to represent the entire DOE height map. The parameterization reduces both the number of learnable parameters and manufacturing complexity.
The height map $H(x,y)$ of a rotationally symmetric DOE can be represented as follows:
\begin{equation}
	H(x,y) = H_{P}(x,y,\theta_{height map}),
\end{equation}	
where the $H_{P}$ is the height map, and the $\theta_{height map}$ is a one-dimensional learnable parameter.

According to diffraction law, the PSF $P(x,y,\lambda)$ can be derivated from the height map $H(x,y)$. For the detailed derivation, 
please refer to the supplementary materials for additional information.
Then, the imaging process of \mbox{PEM-P} can be written based on UEM:
\begin{equation}
	I_{rgb} = \int W(\lambda) \cdot (P(x,y,\lambda) \ast I(x,y,\lambda)) d\lambda.
	\label{pemp_imaging}
\end{equation}

\noindent {\bf{Physical Wavelength Encoding Model.}} 
The \mbox{WEM-P} algorithm selects the best spectral response curve from a database while keeping other system components fixed, ensuring optimal performance for non-learnable spectral response curves~\cite{2020Joint}.
\begin{equation}
	W(\lambda) = \{w_i\vert w_i \in \mathbf{W}, w_i\ is\ the \ most \ suited\},
\end{equation} 
where $w_1,w_2,w_2,... \in \mathbf{W}$ represents a fixed response function dataset.

The imaging process of \mbox{WEM-P} can be written from UEM by eliminating unused terms:
\begin{equation}
	I_{rgb} = \int W(\lambda) \cdot I(x,y,\lambda) d\lambda.
	\label{wemp_imaging}
\end{equation}

\subsection{Ideal Model}
We then introduce the ideal UEMs, \mbox{AEM-I}, \mbox{PEM-I}, and \mbox{WEM-I}, free from any physical constraint in the physical UEMs.

\noindent {\bf{Ideal Amplitude Encoding Model.}} 
Compared to \mbox{AEM-P}, which uses binary spatial dimensional encoding, we design the \mbox{AEM-I} using floating point encoding to encode spatial and spectral dimensional encoding mask $A_{I}(x,y,\lambda)$.

The imaging model of \mbox{AEM-I} is obtained by replacing $A(x,y,\lambda)$ in Eq.~(\ref{aemp_imaging}) with $A_{I}(x,y,\lambda)$.
\begin{equation}
	A_{I}(x,y,\lambda) = Mask_{I}(x,y,\lambda,\theta_{mask}),
\end{equation}	
where the $Mask_{I}$ is the spatial and spectral dimensional encoding mask, the $\theta_{mask}$ is a learnable parameters.

\noindent {\bf{Ideal Phase Encoding Model.}} 
We encode the PSF $P_I(x,y,\lambda)$ directly for the \mbox{PEM-I} without considering the structure of the DOE of \mbox{PEM-P}.
The imaging model of \mbox{PEM-I} is obtained by replacing $P(x,y,\lambda)$ in Eq.~(\ref{pemp_imaging}) with $P_I(x,y,\lambda)$.
\begin{equation}
	P_I(x,y,\lambda) = PSF_I(x,y,\lambda,\theta_{psf}),
\end{equation}	
where the $PSF_{I}$ is the encoding PSF, the $\theta_{psf}$ is a learnable parameters.

\noindent {\bf{Ideal Wavelength Encoding Model.}} 
When designing filter surface coatings, customizing the spectral curve must consider filter material properties.
Unlike \mbox{WEM-P}, our \mbox{WEM-I} encodes the spectral curves directly but does not consider the material limitations.
The imaging model of \mbox{PEM-I} is obtained by replacing $W(x,y,\lambda)$ in Eq.~(\ref{wemp_imaging}) with $W_I(x,y,\lambda)$.
\begin{equation}
	W_{I}(\lambda) = Response_{I}(\lambda,\theta_{response}),
\end{equation}
where the $Response_{I}$ is the encoding response function, and the $\theta_{response}$ is a learnable parameters.

\begin{table*}[!t]
	\centering
	\caption{Spectral imaging performance of systems using UEMs on different datasets, with the best results in bold.}
	\label{table_simconv}
	\resizebox{\linewidth}{!}{
		\begin{tabular}{c|c|cccc|c|cccc}
			\hline
			~ & \multicolumn{5}{c|}{\textbf{Physical Model}}& \multicolumn{5}{c}{\textbf{Ideal Model}} \\ \hline
			\textbf{Datasets} & \textbf{Encoding Model} & \textbf{PSNR$\uparrow$} & \textbf{PSNR-SI$\uparrow$} & \textbf{SAM$\downarrow$} & \textbf{ERGAS$\downarrow$} & \textbf{Encoding Model} & \textbf{PSNR$\uparrow$} & \textbf{PSNR-SI$\uparrow$} & \textbf{SAM$\downarrow$} & \textbf{ERGAS$\downarrow$}  \\ \hline
			\multirow{2}*{\textbf{ICVL}} & \textbf{WEM-P} & 44.96 & 38.81 & 0.0399 & 6.23 & \textbf{WEM-I} & \textbf{45.25} & 38.51 & \textbf{0.0328} & \textbf{5.23}  \\ 
			~ & \textbf{AEM-P} & 40.21 & 32.66 & 0.0491 & 8.96 & \textbf{AEM-I} & 45.08 & \textbf{38.95} & 0.0405 & 6.24  \\ 
			\multirow{2}*{(training \& testing)} & \textbf{PEM-P} & 34.07 & 26.35 & 0.0669 & 16.16 & \textbf{PEM-I} & 44.9 & 38.68 & 0.0398 & 6.28  \\ 
			~ & \textbf{Baseline} & 32.21 & 28.98 & 0.1342 & 33.3 &    &   &   &   &   \\ \hline
			\multirow{2}*{\textbf{Harvard}} & \textbf{WEM-P} & 38.72 & 36.91 & 0.07538 & 8.01 & \textbf{WEM-I} & \textbf{39.9} & 36.31 & \textbf{0.05152} & 8.09  \\ 
			~ & \textbf{AEM-P} & 32.7 & 30.01 & 0.0852 & 14.04 & \textbf{AEM-I} & 38.84 & \textbf{36.88} & 0.07513 & \textbf{7.97}  \\ 
			\multirow{2}*{(training \& testing)} & \textbf{PEM-P} & 27.24 & 24.42 & 0.10196 & 25.17 & \textbf{PEM-I} & 38.71 & \textbf{36.88} & 0.07561 & 8.12  \\ 
			~ & \textbf{Baseline} & 26.28 & 24.87 & 0.21845 & 143.05 &    &   &   &   &   \\ \hline
			\multirow{2}*{\textbf{CAVE}} & \textbf{WEM-P} & 36.02 & 37.57 & 0.1961 & 18.38 & \textbf{WEM-I} & \textbf{37.65} & 37.98 & 0.1682 & 17.48  \\ 
			~ & \textbf{AEM-P} & 33.13 & 33.13 & 0.2193 & 26.11 & \textbf{AEM-I} & 36.96 & \textbf{38.00} & \textbf{0.1590} & \textbf{16.85}   \\ 
			\multirow{2}*{(training \& testing)} & \textbf{PEM-P} & 28.99 & 28.76 & 0.2974 & 48.23 & \textbf{PEM-I} & 35.77 & 37.55 & 0.1894 & 18.77  \\ 
			~ & \textbf{Baseline} & 29.50 & 30.36 & 0.3906 & 41.74 &    &   &   &   &   \\  \hline
			\multirow{2}*{\textbf{CAVE} (training),} & \textbf{WEM-P} & 32.02 & 33.58 & 0.4819 & 36.82 & \textbf{WEM-I} & 33.04 & 31.93 & 0.4363 & 43.45  \\ 
			~ & \textbf{AEM-P} & 32.21 & 33.69 & 0.4707 & 36.22 & \textbf{AEM-I} & \textbf{33.30} & \textbf{34.42} & \textbf{0.3679} & \textbf{31.82}  \\ 
			\multirow{2}*{\textbf{KAIST} (testing)} & \textbf{PEM-P} & 28.39 & 26.20 & 0.4909 & 70.45 & \textbf{PEM-I} & 30.68 & 29.60 & 0.4541 & 48.82  \\ 
			~ & \textbf{Baseline} & 30.78 & 31.06 & 0.6339 & 46.47 &    &   &   &   &   \\ \hline
	\end{tabular}}
\end{table*}

\begin{table*}[!ht]
	\centering
	\caption{Spectral imaging performance of systems using UEMs on different decoding models, with the best results in bold.}
	\label{table_decodingmodel}
	\resizebox{\linewidth}{!}{
		\begin{tabular}{c|c|cccc|c|cccc}
			\hline
			~ & \multicolumn{5}{c|}{\textbf{Physical Model}}& \multicolumn{5}{c}{\textbf{Ideal Model}} \\ \hline
			\textbf{Decoding Model} & \textbf{Encoding Model} & \textbf{PSNR$\uparrow$} & \textbf{PSNR-SI$\uparrow$} & \textbf{SAM$\downarrow$} & \textbf{ERGAS$\downarrow$} & \textbf{Encoding Model} & \textbf{PSNR$\uparrow$} & \textbf{PSNR-SI$\uparrow$} & \textbf{SAM$\downarrow$} & \textbf{ERGAS$\downarrow$}  \\ \hline
			\multirow{3}*~ & \textbf{WEM-P} & 38.62 & 32.00 & 8.07 & 0.0373 & \textbf{WEM-I} & \textbf{40.54} & \textbf{33.96} & \textbf{6.73} & \textbf{0.0360}  \\ 
			\textbf{Res-U-Net} & \textbf{AEM-P} & 36.72 & 29.53 & 10.39 & 0.0422 & \textbf{AEM-I} & 39.28 & 32.80 & 7.51 & 0.0396  \\ 
			~ & \textbf{PEM-P} & 34.87 & 28.10 & 13.08 & 0.0654 & \textbf{PEM-I} & 38.57 & 31.83 & 8.15 & 0.0385  \\ \hline
			\multirow{3}*~ & \textbf{WEM-P} & 45.09 & 38.93 & 4.22 & 0.0289 & \textbf{WEM-I} & \textbf{46.43} & \textbf{40.86} & \textbf{3.69} & \textbf{0.0257}  \\ 
			\textbf{Unfolding-Net} & \textbf{AEM-P} & 44.63 & 38.55 & 4.43 & 0.0295 & \textbf{AEM-I} & 46.33 & 40.30 & 3.71 & 0.0280  \\ 
			~ & \textbf{PEM-P} & 38.70 & 31.80 & 8.69 & 0.0514 & \textbf{PEM-I} & 45.94 & 40.53 & 3.84 & 0.0286  \\ \hline
	\end{tabular}}
\end{table*}

\section{Joint Optimization}
This section introduces the decoding models we use and the optimization objective in the joint optimization framework for the comparative study.

\noindent {\bf{Decoding Model.}} 
Since we mainly explore the performance of several encoding models rather than the decoding models, we design three typical reconstruction networks as decoding models. These include a Sim-Conv-Net which is a simple CNN~\cite{lecun1998gradient,lee2017going,maffei2019single}, Res-U-Net~\cite{zhang2018road,gurrola2021residual} which is a variant of the popular U-Net~\cite{ronneberger2015u,miao2019net} and Unfolding-Net which uses deep unfolding methods~\cite{dong2018denoising,zhang2020deep,zhang2020amp,sun2021deep,zhang2021plug,zhang2022herosnet,cai2022degradation} as the reconstruction network. For the detailed network structure, please refer to the supplementary materials. 

\noindent {\bf{Joint Optimization Framework.}} 
We combine the UEM with the decoding model to compose a joint encoder-decoder optimization framework.
In UEM, the scene light passes through the mask to get the amplitude modulation $f_A(I,\theta_{A})$, passes through the DOE to get the phase modulation $f_P(I,\theta_{P})$, passes through the filter and sensor to get the wavelength modulation$f_W(I,\theta_{W})$. We can describe the overall imaging process as follows:
\begin{equation}
	I_{rgb} = f_W(f_P(f_A(I,\theta_{A}),\theta_{P}),\theta_{W}),
\end{equation}		
where $\theta_{A}$,$\theta_{P}$, and $\theta_{W}$ are learnable parameters for the AEM, the PEM, and the WEM. For each system only optimizes one parameter, e.g., for AEM systems, only $\theta_{A}$ is optimized. After that, the decoding model reconstructs the spectral images from the RGB measurement $I_{rgb}$, and this process can be written as:
\begin{equation}
	\tilde{I} = f_{D}(I_{rgb},\theta_{D}),
\end{equation}	
where the $\theta_{D}$ is the learnable parameter of the decoding model.

Then, the optimization objective of the UEM is to find a set of $\theta_{A},\theta_{P},\theta_{W},\theta_{D}$ that minimizes the mean absolute error (MAE) loss between the reconstructed spectral images and the ground truth:
\begin{equation}
	\theta_{A},\theta_{P},\theta_{W},\theta_{D} = \mathop{\arg\min}_{\theta_{A},\theta_{P},\theta_{W},\theta_{D}}{\sum \Vert\tilde{I} - I\Vert}_1.
\end{equation}

Since each stage of the UEM is differentiable, we can solve this problem using the first-order gradient-based optimization methods. The irrelevant parameter will be eliminated when we cast the UEM to a specific encoding model. The irrelevant parameter will be eliminated. That is, we only optimize $\theta_{A},\theta_{D}$ for AEMs, $\theta_{P},\theta_{D}$ for PEMs, and $\theta_{W},\theta_{D}$ for WEMs.

Finally, we combine the UEM, decoding model, and optimization objective to build a joint encoder-decoder optimization framework, as shown in Figure~\ref{fig_joint_optimization_framework}.

\section{Experiments and Analysis}

\subsection{Experimental Setup}

In our experiments, the Sim-Conv-Net serves as the primary decoding model, and we establish a baseline by employing nearest neighbor interpolation from RGB images to the spectral image. The camera response curve of the FLIR BFS\_U3\_04S2C\_C is utilized in experiments where the response is fixed (\mbox{AEM-P/I}, \mbox{PEM-P/I}, and \mbox{WEM-P}).

Our joint encoder-decoder optimization framework undergoes evaluation on four datasets: ICVL\cite{arad2016sparse}, Harvard~\cite{chakrabarti2011statistics}, CAVE~\cite{CAVE_0293}, and KAIST~\cite{DeepCASSI:SIGA:2017}, with cross-validation. 
The dataset sizes are as follows: ICVL has 167 training, 17 validation, and 17 testing images. Harvard has 40 training, 4 validation, and 6 testing images. CAVE has 28 training, 1 validation, and 3 testing images. Lastly, KAIST has 30 testing images exclusively.
All images are cropped to 512 × 512 patches. 

We consider PSNR, PSNR-SI, SAM, and ERGAS for evaluation metrics. The supplementary material provides more details on these metrics.

\begin{figure}[!t]
	
	\centering
	\subfloat[GT]{\begin{overpic}[width=0.7in]{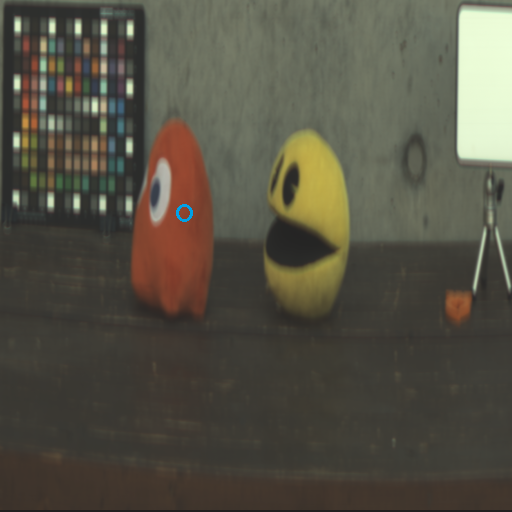}
	\end{overpic}\label{rec_GT}}\label{1}
	\subfloat[WEM-I]{\begin{overpic}[width=0.7in]{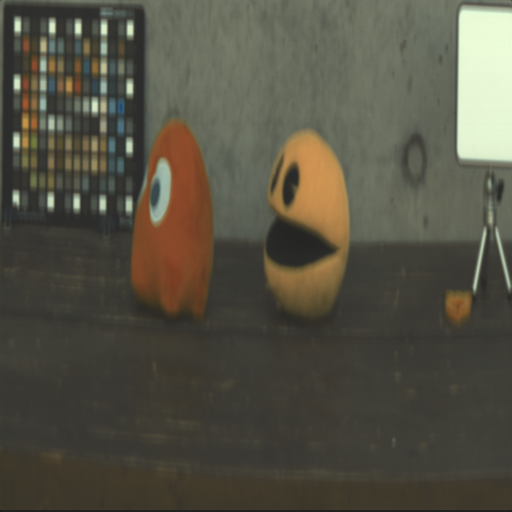}
                \put(43,5){\footnotesize \color{white} 41.97dB}
        \end{overpic}\label{reconstructed_coded_sensor}}\label{1}
	\subfloat[AEM-I]{\begin{overpic}[width=0.7in]{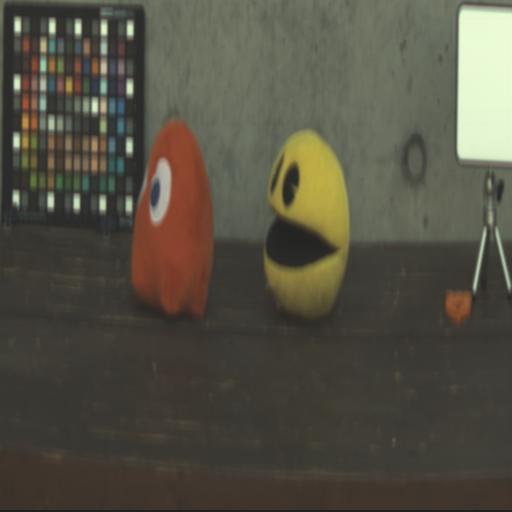}
			\put(43,5){\footnotesize \color{white} 41.24dB }
	\end{overpic}}\label{1}
	\subfloat[PEM-I]{\begin{overpic}[width=0.7in]{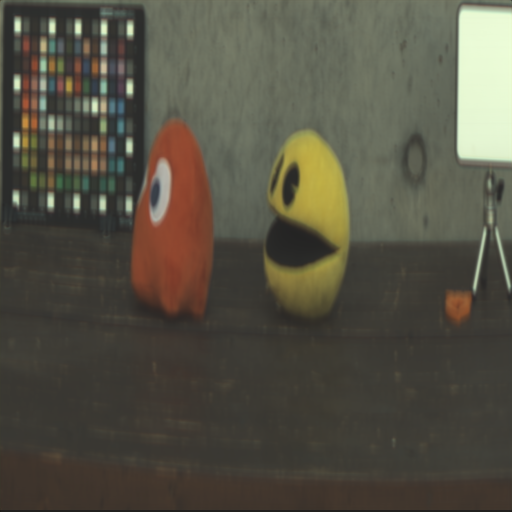}
			\put(43,5){\footnotesize \color{white} 40.99dB }
	\end{overpic} }
	
	\subfloat[Baseline]{\begin{overpic}[width=0.7in]{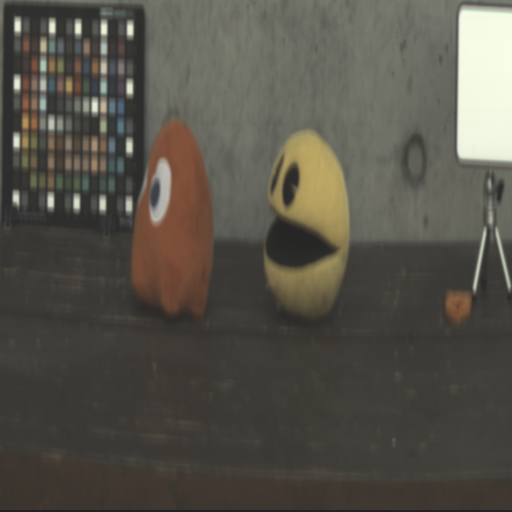}
			\put(43,5){\footnotesize \color{white} 34.58dB }
	\end{overpic}}\label{1}
	\subfloat[WEM-P]{\begin{overpic}[width=0.7in]{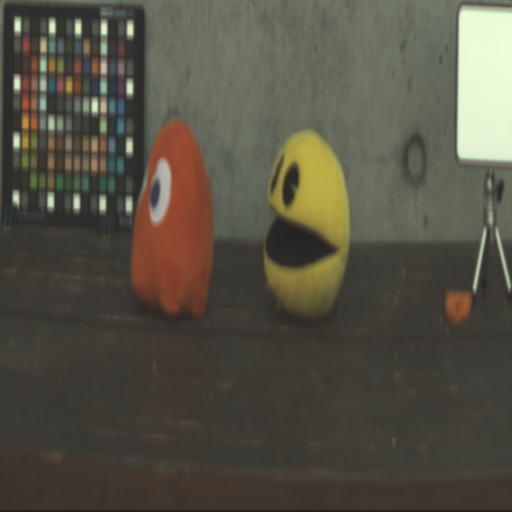}
			\put(43,5){\footnotesize \color{white} 41.79dB }
	\end{overpic}}\label{1}
	\subfloat[AEM-P]{\begin{overpic}[width=0.7in]{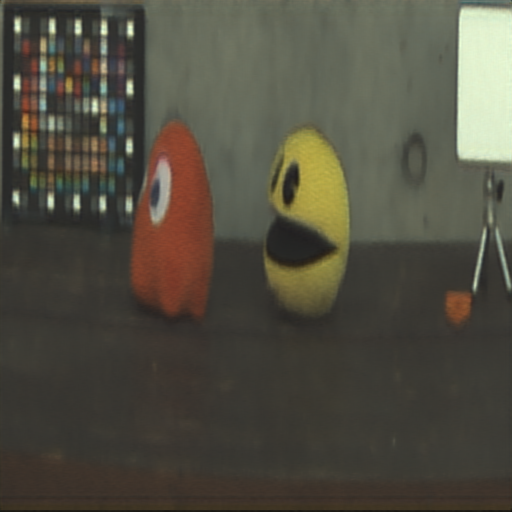}
			\put(43,5){\footnotesize \color{white} 40.75dB }
	\end{overpic}}\label{1}
	\subfloat[PEM-P]{\begin{overpic}[width=0.7in]{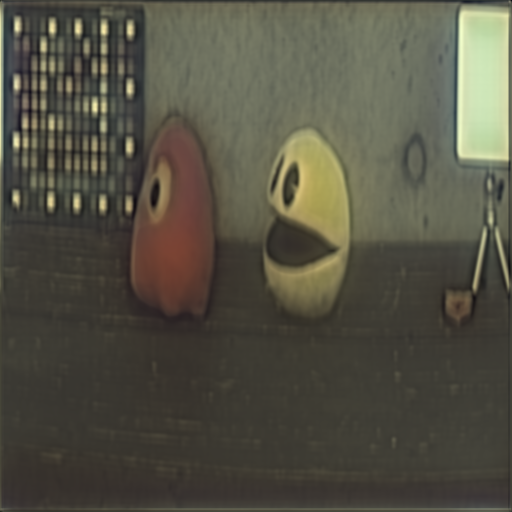}
			\put(43,5){\footnotesize \color{white} 33.05dB }
	\end{overpic}}
	
	\caption{Visual comparison of reconstruction images, the PSNR metrics are labeled in the lower left corner.}
	\label{fig_reconsted_result}
\end{figure}

\begin{figure}[!t]
	\centering
	{\includegraphics[width=3in]{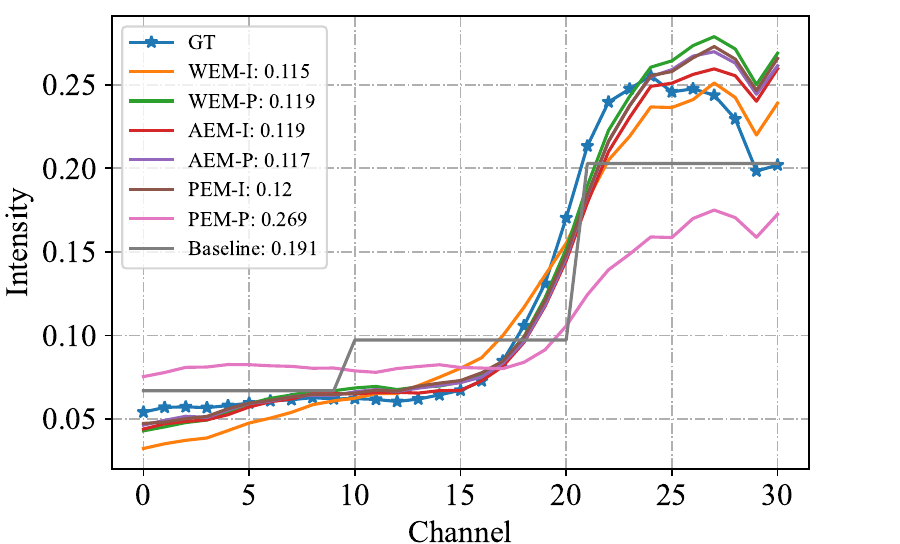}}
	\caption{Spectral curve comparison at a single sampling point, the SAM metrics for this point are labeled in the legend.}
	\label{fig_red_curve}
\end{figure}

\subsection{Results}

\noindent {\bf{Comparative Experiments on Different Datasets.}} 
Table~\ref{table_simconv} presents a comprehensive comparison of the spectral imaging performance of the system, incorporating UEMs and Sim-Conv-Net, across different datasets. The imaging effects of various UEM sub-models are carefully examined, revealing consistent regularity. \mbox{WEM-P} outperforms \mbox{AEM-P} among physical models, while \mbox{PEM-P} exhibits relatively inferior performance.
Similarly, in the ideal model, \mbox{WEM-I} imaging is superior. Given that \mbox{WEM-P} already has the best performance in the physical model, the \mbox{WEM-I} brings relatively little improvement. In contrast, \mbox{AEM-P} and \mbox{PEM-P} show great potential for improvement when transitioning to the ideal model.

To evaluate the generalizability of our systems, we train on the CAVE dataset and test on the KAIST dataset, using only overlapping bands for both training and testing. Significantly, AEMs demonstrate superior generalizability, whereas \mbox{WEM-P} and \mbox{PEM-I}, when trained on CAVE, exhibit noticeable performance dips when tests on KAIST. 
Additional validation experiments on ICVL are detailed in the supplementary materials.

RGB visualizations of the reconstructed spectral images for all models are displayed in Figure~\ref{fig_reconsted_result}. Although \mbox{WEM-I} performs better numerically, we notice slight color inaccuracies in some instances, such as a dominant orange shift (Figure~\ref{reconstructed_coded_sensor}). We attribute this shift to the negative values in the response curve of the free encoding. \mbox{PEM-P} consistently shows inferior reconstruction, as reflected in color and image detail, in line with numerical metrics.

In addition to the numerical and visual evaluations, we select a point in Figure~\ref{fig_reconsted_result} (marked in the blue-circled area in Figure~\ref{rec_GT}) to obtain the spectral accuracy of each model for comparing the spectral accuracies of the models, as shown in Figure~\ref{fig_red_curve}. 
The results indicate that the \mbox{WEM-I} model curve aligns with the ground truth most precisely, while the \mbox{PEM-P} model is less accurate than the baseline. The other physical and idea models display relatively close curves around the ground truth, consistent with the numeric metrics.

\noindent {\bf{Experiments on Different Decoding Model.}} 
In this experiment, we replace the Sim-Conv-Net decoding model with two different reconstruction networks: the Res-U-Net and the Unfolding-Net. Table~\ref{table_decodingmodel} summarizes the results.

Contrary to expectations that increasing network depth and complexity would enhance image reconstruction, Res-U-Net results in lower metrics in our end-to-end optimization experiments. Except for \mbox{WEM-I}, the tables display those models utilizing Sim-Conv-Net have higher metrics sensitive to spatial differences like PSNR. In contrast, models using Res-U-Net as the decoding model tend to have superior metrics sensitive to spectral errors like SAM. Only \mbox{PEM-P} shows enhanced imaging performance when applying Res-U-Net compared to Sim-Conv-Net.

Unfolding-Net incorporates optical encoding information into the inference process and achieves superior decoding performance by optimizing model-driven and data-driven modules iteratively. Compared to Sim-Conv-Net, physical model systems demonstrate significantly improved performance, while ideal model systems exhibit only marginal gains.

Training a large network structure in an end-to-end optimized computational spectral imaging system may be challenging. Thus, we suggest utilizing simpler convolutional neural networks, particularly in situations with limited computing power. Moreover, we advise that simpler optical encoding processes are more compatible with simple networks, while more complex encoding models should prefer deeper networks.

\begin{table}[!t]
	\centering
	\caption{Comparison of the segmentation quality of all systems, with the best results in bold and the next best underlined.}
	\label{table_segmentation}
	\resizebox{\linewidth}{!}{
		\begin{tabular}{cc|cc}
			\hline
			\multicolumn{2}{c|}{\textbf{Physical Model}}& \multicolumn{2}{c}{\textbf{Ideal Model}} \\ \hline
			\textbf{Encoding Model} & \textbf{Accuracy$\uparrow$} & \textbf{Encoding Model} & \textbf{Accuracy$\uparrow$} \\ \hline
			\textbf{WEM-P} & 0.5218 & \textbf{WEM-I} & \underline{0.5221}\\ 
			\textbf{AEM-P} & 0.4151 & \textbf{AEM-I} & \textbf{0.5234}\\ 
			\textbf{PEM-P} & 0.4479 & \textbf{PEM-I} & 0.5128 \\
			\textbf{Baseline} & 0.3223 &   &   \\\hline
	\end{tabular}}
\end{table}

\begin{table}[!t]
	\centering
	\caption{Comparison of the imaging quality of systems using different types of WEM, where the WEM w/ P.C. represents the wavelength encoding model with positive constraints. The best result is in bold, and the next best is underlined.}
	\label{table_codedsensor}
	\resizebox{\linewidth}{!}
	{\begin{tabular}{ccccc}
			\hline
			\textbf{Encoding Model} & \textbf{PSNR$\uparrow$} & \textbf{PSNR-SI$\uparrow$} & \textbf{SAM$\downarrow$} & \textbf{ERGAS$\downarrow$} \\ \hline
			\textbf{WEM-I} & \textbf{45.25} & 38.51 & \textbf{0.0328} & \textbf{5.23} \\ 
			\textbf{WEM-P} & \underline{44.96} & \underline{38.81} & \underline{0.0399} & 6.23 \\ 
			\textbf{WEM-I w/ P.C.} & 44.26 & \textbf{38.84} & 0.0432 & \underline{5.37} \\ \hline
			
	\end{tabular}}
\end{table}

\noindent {\bf{Experiments on Image Segmentation.}} 
In practical applications, spectral imaging serves various high-level tasks. 
Therefore, evaluating these systems based solely on imaging performance may be insufficient, prompting further exploration into how spectral imaging can aid perceptual tasks.
To evaluate this aspect, we choose spectral image segmentation as a representative high-level task and employ a Res-U-Net as the segmentation network, assessing segmentation performance on the LIB-HSI dataset~\cite{habili2023hyperspectral}.

\begin{figure}[!t]
	\centering
	\subfloat[RGB Image]{\includegraphics[width=0.95in,angle=-90]{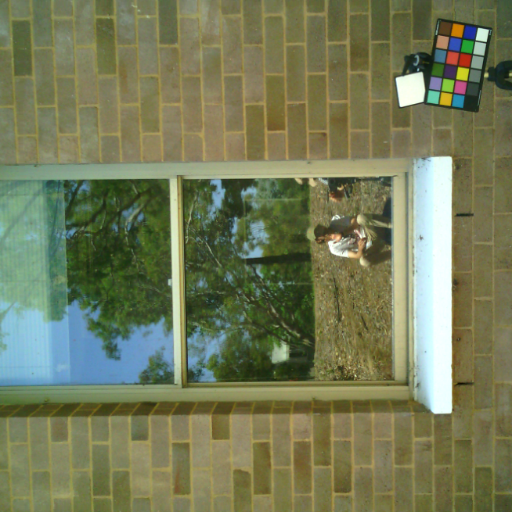}}
	\label{seg_image}
	\subfloat[GT]{\includegraphics[width=0.95in,angle=-90]{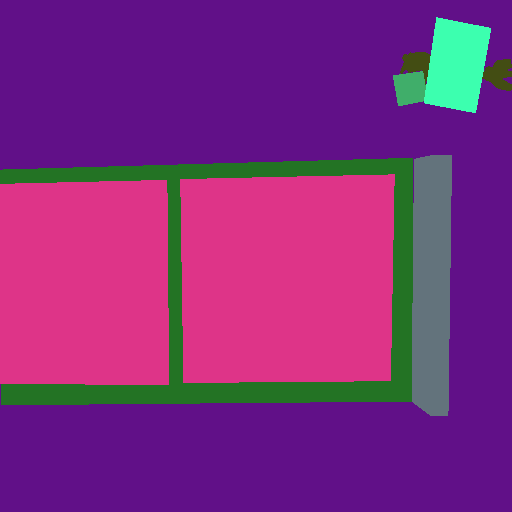}}
	\label{seg_GT}
	\subfloat[Baseline]{\includegraphics[width=0.95in,angle=-90]{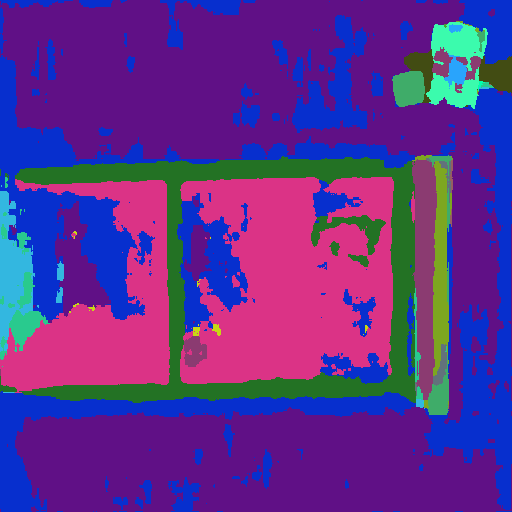}}
	\label{seg_baseline}
	
	\subfloat[WEM-I]{\includegraphics[width=0.95in,angle=-90]{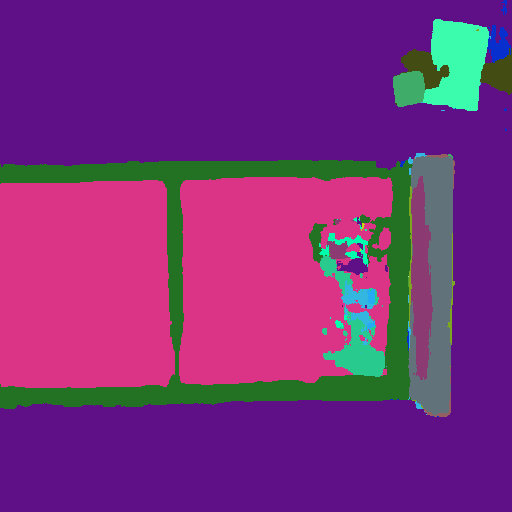}}
	\label{seg_wemi}
	\subfloat[AEM-I]{\includegraphics[width=0.95in,angle=-90]{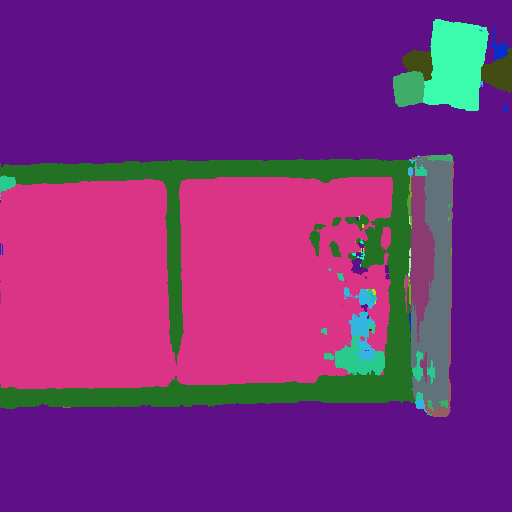}}
	\label{seg_aemi}
	\subfloat[PEM-I]{\includegraphics[width=0.95in,angle=-90]{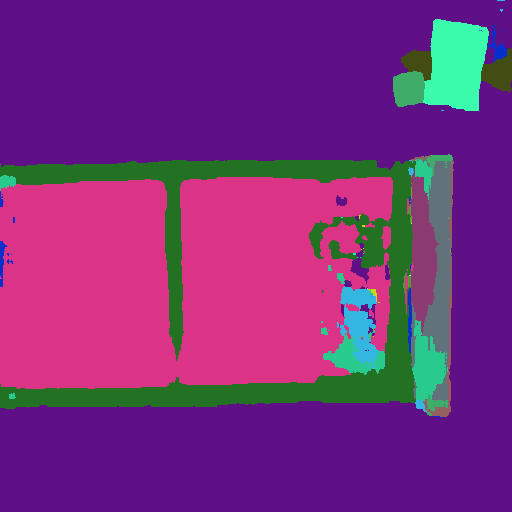}}
	\label{seg_pemi}
	
	\subfloat[WEM-P]{\includegraphics[width=0.95in,angle=-90]{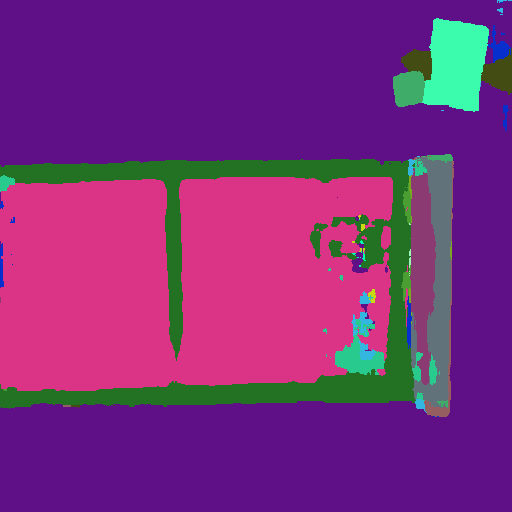}}
	\label{seg_wemp}
	\subfloat[AEM-P]{\includegraphics[width=0.95in,angle=-90]{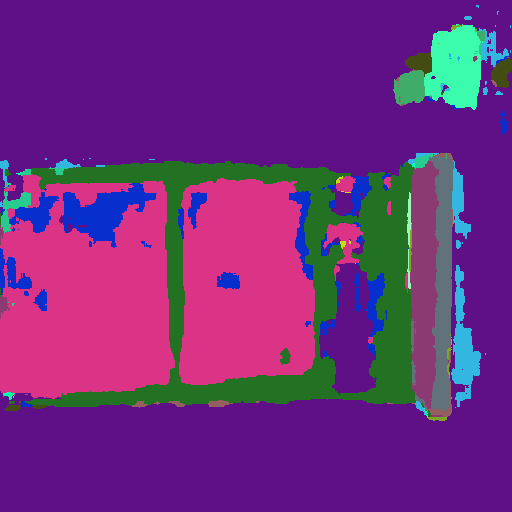}}
	\label{seg_aemp}
	\subfloat[PEM-P]{\includegraphics[width=0.95in,angle=-90]{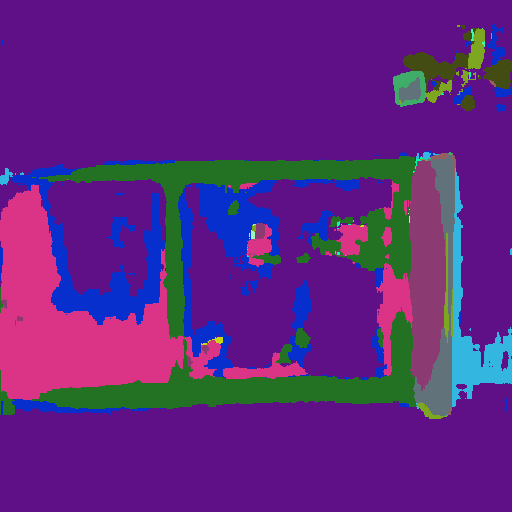}}
	\label{seg_pemp}
	
	\caption{Visual comparison of segmentation results.}
	\label{fig_seg_result}
\end{figure}

Spectral images from each imaging system are fed into the segmentation network for evaluation. Using the same segmentation network, Table~\ref{table_segmentation} compares the segmentation accuracy among the previously discussed UEMs. Figure~\ref{fig_seg_result} shows the visualization results. Although noteworthy, \mbox{AEM-P} outperforms \mbox{PEM-P} in low-level spectral imaging but falls behind in segmentation performance. 
A positive correlation between low-level imaging and high-level segmentation performance cannot be established.
We conclude that phase encoding presents more difficulty decoding overlapping information than amplitude and wavelength encoding. However, phase encoding may retain more comprehensive semantic information about the image, leading to improved performance in image segmentation.


\subsection{Breakdown Analysis}
In this section, we analyze the characteristics of each sub-model of UEM in combination with the experimental results and visualization.

\begin{figure*}[!t]
	\centering
	\subfloat[Sim-Conv-Net, channel 6-11]{\includegraphics[width=3.25in]{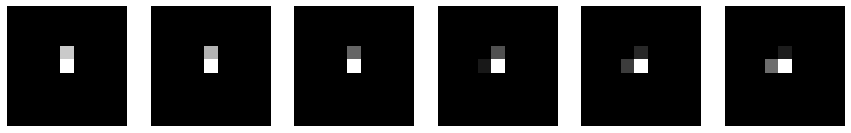}}
	\label{fig_freepsf_sim_6-11}
	\hfil
	\subfloat[Sim-Conv-Net, channel 18-23]{\includegraphics[width=3.25in]{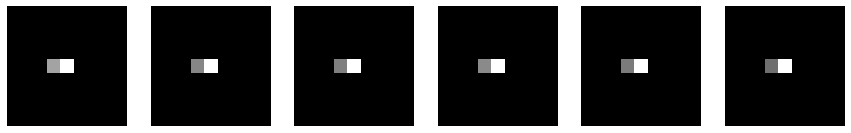}}
	\label{fig_freepsf_sim_18-23}
	
	\subfloat[Res-U-Net (depth=7), channel 6-11]{\includegraphics[width=3.25in]{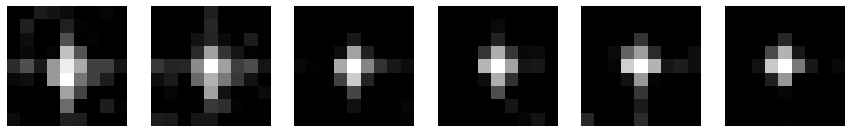}}
	\label{fig_freepsf_res7_6-11}
	\hfil
	\subfloat[Res-U-Net (depth=7), channel 18-23]{\includegraphics[width=3.25in]{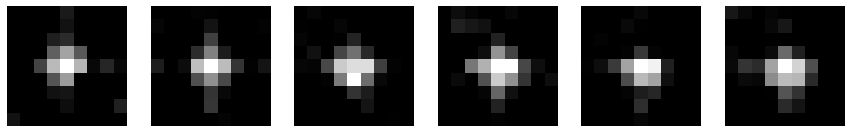}}
	\label{fig_freepsf_res7_18-23}
	
	\caption{Visual comparison of the PSFs of \mbox{PEM-I} using different decoding networks. The PSF energy of the Sim-Conv-Net is more convergent at the center, and the PSF pattern is differentiated between channels. The PSFs of the Res-U-Net are more diffuse and have no obvious regularity.}
	\label{fig_freepsf}
\end{figure*}

\begin{figure*}[!t]
	\centering
	\subfloat[Mask]{\includegraphics[width=2.25in]{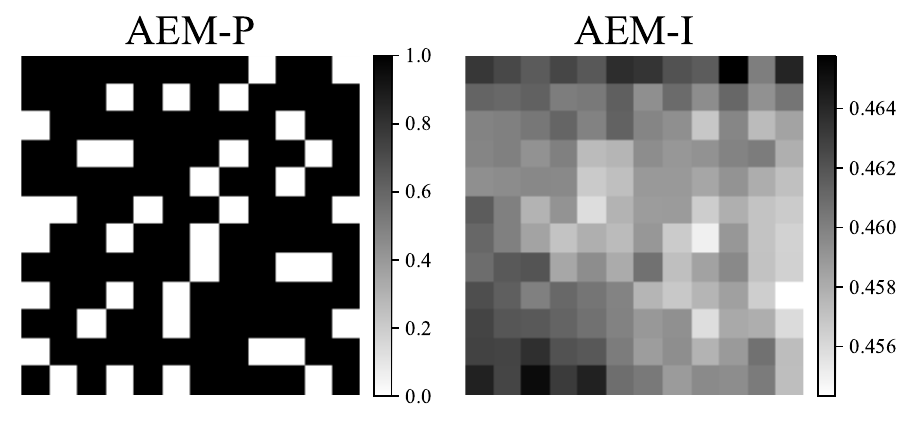}\label{Mask} } \quad
	\subfloat[Filter Response Curve]{\includegraphics[width=4.1in]{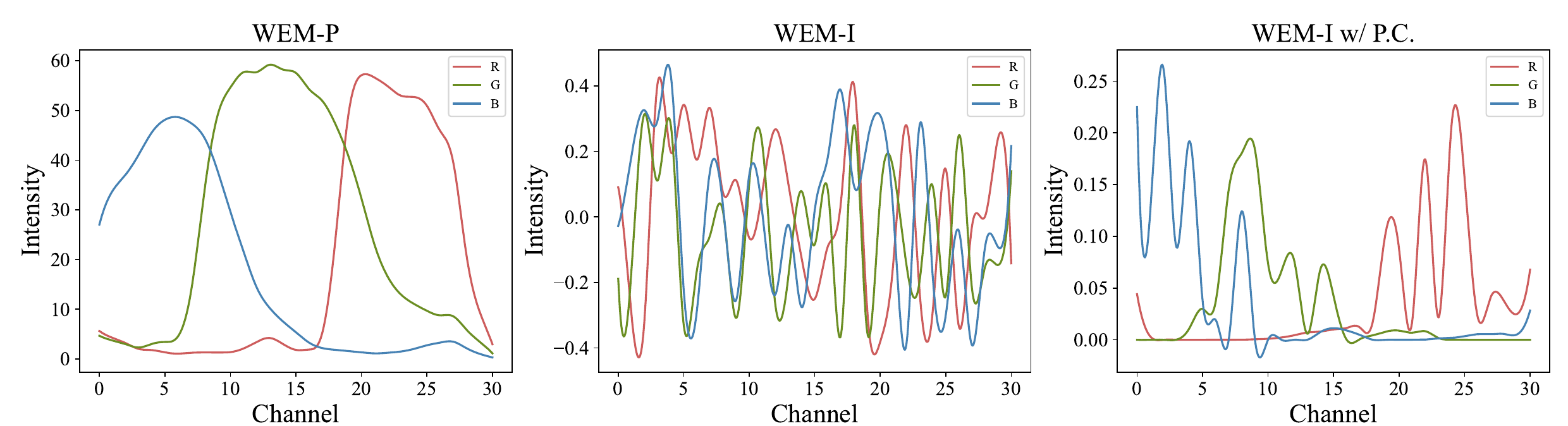}\label{response_fixed}}
	
	\caption{Visual comparison of masks of AEMs and response curves of WEMs. The mask is zoomed in for better visualization.}
	\label{fig_UEMA_MASK}
\end{figure*}

\noindent {\bf{Amplitude Encoding Model.}} 
We show the masks in Figure~\ref{response_fixed}. The \mbox{AEM-I} mask has a narrower value range than the binary \mbox{AEM-P} mask, with clear distinctions between channels, illustrating similar masks amongst neighboring channels. Supplementary material provides a complete view of \mbox{AEM-I} masks.

Retaining more scene information in optical encoding seems advantageous for obtaining superior spectral reconstruction and enhancing robustness in the context of AEMs. Our experimental findings strongly indicate that optimizing AEM masks via a floating-point representation rather than a binary one contributes to boosting the performance of spectral imaging.

\noindent {\bf{Phase Encoding Model.}} 
In \mbox{PEM-I}, achieving convergent optimization results becomes challenging when the PSF has the same size as the image. Consequently, we set the PSF size to 9, optimizing only the central 9 × 9 region and padding the remaining area with zeros. 
The visualization of \mbox{PEM-I} PSFs, utilizing two different decoding models, is depicted in Figure~\ref{fig_freepsf}. Notably, the PSFs of \mbox{PEM-I} with Res-U-Net exhibit significant divergence and inferior imaging performance. Conversely, the Sim-Conv-Net experiment, which attains the best performance, showcases relatively concentrated patterns at the center with slight variations across channels.

Our experiments involve varying PSF sizes and convolutional kernel sizes of the decoding model, revealing that smaller learnable PSF and kernel sizes improve PSF convergence and overall performance. Additional details can be found in the supplementary materials. 
To achieve clear imaging quality and preserve some spectral information, the design of a concentrated PSF with a slight variance across the channels is necessary for the PEM.




\noindent {\bf{Wavelength Encoding Model.}} 
In \mbox{WEM-I}, a linear layer is employed for encoding instead of a fixed RGB filter response function. We visualize the weights of the linear layer as spectral response curves for comparison with the FLIR BFS-U3-04S2C-C. As shown in Figure~\ref{response_fixed}, the response curve of the \mbox{WEM-I} is densely serrated, while the \mbox{WEM-I} with positive constraints  (\mbox{WEM-I} w/ P.C.) is smoother but still has more peaks than the ordinary camera response curve. Comparing the performance of the three WEMs in Table~\ref{table_codedsensor}, the \mbox{WEM-I} still has the best imaging performance, while the \mbox{WEM-I} w/ P.C. and the \mbox{WEM-P} exhibit comparable imaging performance.

The filter and sensor response is crucial in controlling the information bottleneck between the encoder and decoder. When the response is optimized according to the task, the response curve will usually have multiple peaks. According to our results, the ideal shape of the filter response curve tends to have various separated sharp peaks compared to the ordinary response curves of cameras. Therefore, we recommend maximizing peak separation in the design process to improve spectral imaging performance.

\section{Conclusion}

In this study, we propose the UEM, which allows for a fair comparison of amplitude, phase, and wavelength encoding computational spectral imaging systems through joint encoder-decoder optimization. We utilize the UEM to establish physical models of these systems, enabling a comprehensive performance analysis. Additionally, we extend the UEM to ideal models and explore the potential upper bounds of performance without physical constraints. 
We also evaluate these systems on the spectral image segmentation task to investigate the correlation between low-level and high-level task performance. 
Finally, by visualizing and analyzing the ideal UEMs, we provide suggestions and insights for the future design of spectral imaging systems. 
{
    \small
    \bibliographystyle{ieeenat_fullname}
    \bibliography{main}
}


\end{document}


\maketitle
%

\appendix

Due to the space limit, some formula derivation and additional experiments could not be included in the main paper.
This supplementary material provides a comprehensive demonstration of the experimental setup and comparative analysis, structured as follows:
\begin{enumerate}[leftmargin=0.85cm]
	\item[\ref{sec:psf}] PSF Derivation
	\item[\ref{sec:dm}] Decoding Models
	\begin{enumerate}
		\item[\ref{sec:dm_struct}] Structure of Decoding Models
		\item[\ref{sec:dm_uf}] Deep Unfolding Algorithm
	\end{enumerate}
	\item[\ref{sec:ae}] Additional Experiments
	\begin{enumerate}
		\item[\ref{sec:ad_response_selection}] Response Selection of WEM-P
		\item[\ref{sec:ad_init}] Initialization of Ideal Encoding Models
		\item[\ref{sec:ae_loss}] Selection of Loss Functions
		\item[\ref{sec:ae_kernel}] PSF Sizes and Kernel Sizes of PEM-I
		\item[\ref{sec:ae_uem}] Full Optimization 
		\item[\ref{sec:ae_icvl}] Validation of ICVL dataset
	\end{enumerate}
	\item[\ref{sec:v}] Visualization of Ideal Encoding Patterns
\end{enumerate}

\section{PSF Derivation}\label{sec:psf}

Here, we describe the details of the PSF derivation for PEM-P by following the general formulation in~\cite{sitzmann2018end,sun2020learning,metzler2020deep,chang2019deep,baek2021single,li2022quantization}. 

Scene light generates phase delays $\phi(x,y,\lambda)$ through the DOE of the height map $H(x,y)$:
\begin{equation}
	\phi(x,y,\lambda) = \frac{2\pi \Delta n}{\lambda}H(x,y)
\end{equation}	                                  
where $\Delta n$ is the refractive index difference between the air and the material of the optical element.

The incident point source with coordinates $(x,y)$ at a distance d from the DOE can be expressed as:
\begin{equation}
	U_0(x,y,\lambda) = e^{i \frac{2\pi}{\lambda} \frac{x^2+y^2}{d}}
\end{equation}	

The incident light gets phase modulated through the DOE:
\begin{equation}
	U_{doe}(x,y,\lambda) = A(x,y)U_0(x,y,\lambda)e^{i \frac{2\pi}{\lambda} \phi(x,y,\lambda) }
\end{equation}	
where $A(x,y)$ is the aperture optical aperture.

The modulated wavefield passes through the Fresnel diffraction law to reach the sensor plane at a distance $z$ from the DOE:
\begin{small}
	\begin{equation}
		U_{sensor}(x,y,\lambda) =\mathcal{F}^{-1}\left\{\mathcal{F}\left\{U_{doe}(x,y,\lambda)\right\} e^{i \frac{2 \pi}{\lambda} z} e^{-i \pi \lambda z\left(f_x^2+f_y^2\right)}\right\}
	\end{equation}	
\end{small} 

\noindent where $f_x$ and $f_y$ are the frequency variables of $x$ and $y$, respectively, and $\mathcal{F}$ denotes the Fourier transform.

The PSF $P(x, y, \lambda)$ is the squared intensity of the wavefield:
\begin{equation}
	P(x, y, \lambda) \propto\left|U_{sensor}(x,y,\lambda)\right|^2
\end{equation}

\begin{figure*}[!t]
	\centering
	\subfloat[Sim-Conv-Net]{\includegraphics[width=1.5in]{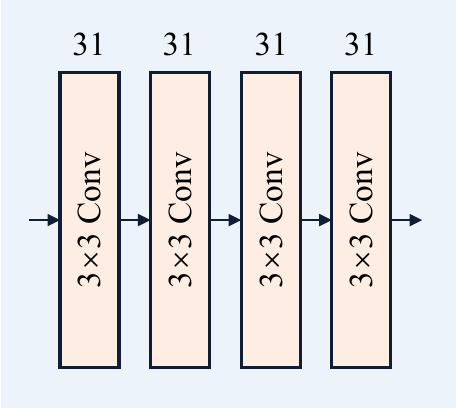}\label{Sim-Conv-Net} } \quad
	\subfloat[Res-U-Net]{\includegraphics[width=4.8in]{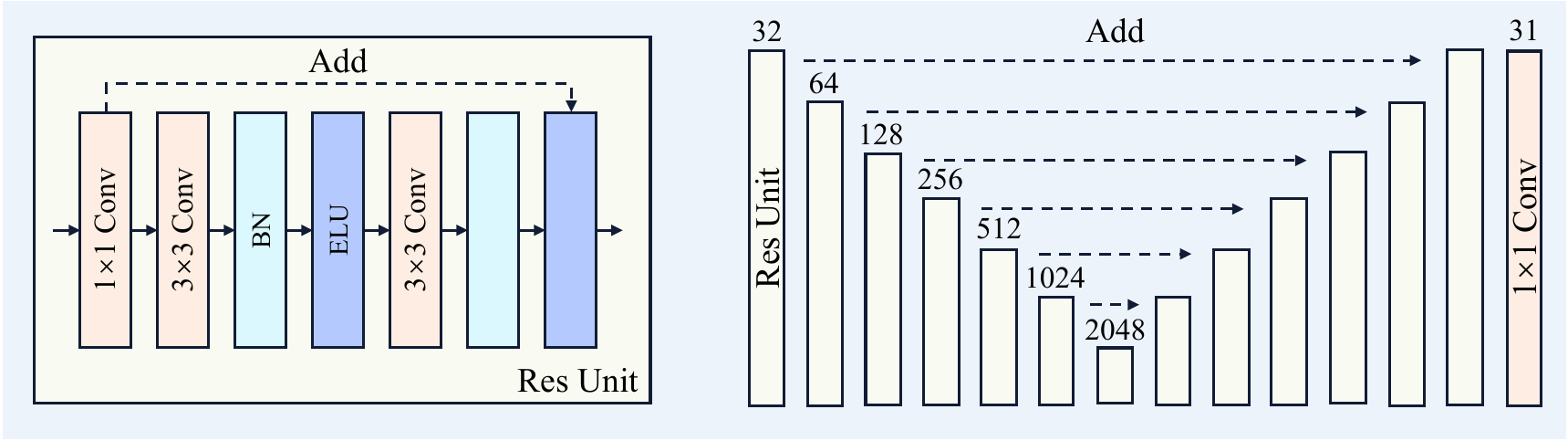}\label{resunet7} }
	
	\subfloat[Unfolding-Net]{\includegraphics[width=4.8in]{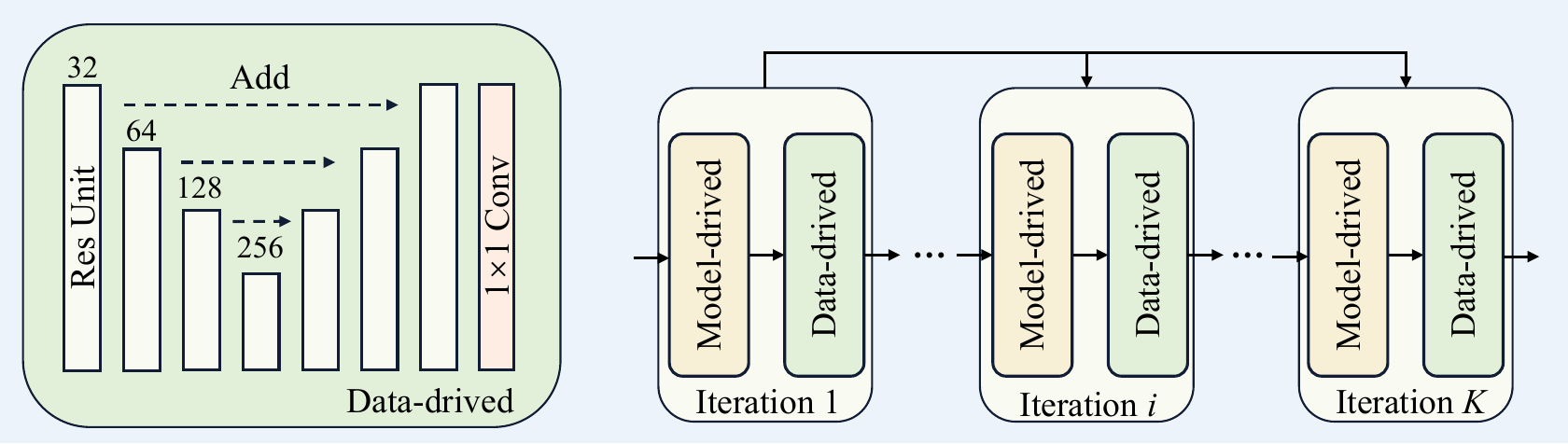}\label{unfolding} }
	
	\caption{The structure of decoding models.}
	\label{fig_decodingmodel}
\end{figure*}

\begin{algorithm}[!t]
	\renewcommand{\algorithmicrequire}{\textbf{Input:}}
	\renewcommand{\algorithmicensure}{\textbf{Output:}}
	\caption{Deep Unfolding Algorithm}
	\label{alg:deepunfolding}
	\begin{algorithmic}[1]
		\REQUIRE {Sensor image $I_{rgb}$, optical encoding operation $\mathcal{O}$, trade-off parameter $\eta$}
		\ENSURE {Reconstructed spectral image $Z_{K}$   }    
		\STATE Initialize $Z_0$ from $I_{rgb}$          
		\FOR {$k = 1, 2, ..., K$ }                  
		\STATE{ $I_k = \arg \min _{I}\|I_{rgb}-\mathcal{O}(I)\|^2+\eta\left\|I-Z_{k-1}\right\|^2 ;$ \\ \textcolor{gray}{//Solving the encoding Model-drived sub-problem}}\\
		\STATE{ $	Z_k= f\left(I_k\right) ;$ \\ \textcolor{gray}{//Reconstructing with a Res-U-Net(depth=4)}}
		\ENDFOR
	\end{algorithmic}
\end{algorithm}

\section{Decoding Models} \label{sec:dm}    
\subsection{Structure of Decoding Models}  \label{sec:dm_struct}

\noindent {\bf{Sim-Conv-Net.}} 
The Sim-Conv-Net, shown in Figure~\ref{Sim-Conv-Net}, comprises four convolutional layers, and the filter size of each convolutional kernel is 31. 

\noindent {\bf{Res-U-Net.}} 
The Res-U-Net,  shown in Figure~\ref{resunet7}, consists of Res Units, the filter size indicated by the top number. Res-U-Net integrates these concepts by incorporating residual blocks into the U-Net architecture. This combination aims to leverage the benefits of both residual connections for gradient flow and U-Net for precise segmentation.

\noindent {\bf{Unfolding-Net.}} 
The Unfolding-Net, shown in Figure~\ref{unfolding}, is based on the deep unfolding algorithm~\cite{zhang2017learning, dong2018denoising, wang2019hyperspectral, zhang2020deep, zhang2021plug} and consists of a model-driven module and a data-driven module. The algorithmic process is shown in Algorithm~\ref{alg:deepunfolding}, and the final reconstruction result is obtained by iterating these two modules several times. In the experiment, the number of iterations $K$ is 4.

\subsection{Deep Unfolding Algorithm} \label{sec:dm_uf}

In general, spectral image reconstruction aims at recovering the potential spectral image $I$ from the RGB sensor observation $I_{rgb}= \mathcal{O}(I)+n$, where $\mathcal{O}$ is a noise-independent optical encoding operation and $n$ is assumed to be additive white Gaussian noise with a standard deviation of $\sigma$.

From a Bayesian perspective~\cite{bayes1958essay}, the solution $\hat{I}$ can be obtained by solving a Maximum A Posteriori (MAP) estimation problem.

\begin{equation}
	\hat{I}=\underset{I}{\arg \max } \log p(I_{rgb} \mid I)+\log p(I) 
	\label{map_eq}
\end{equation}
where $\log p(I_{rgb} \mid I)$ represents the log-likelihood of observation $I_{rgb}$, $\log p(I)$ delivers the prior of clean image I and is independent of degraded image $I_{rgb}$. Eq.~\ref{map_eq} can be reformulated as

\begin{equation}
	\hat{I} = argmin{_I} { \lvert\lvert{I_{rgb}-\mathcal{O}(I)}\rvert\rvert^2+\lambda R(I)}
	\label{solve_eq}
\end{equation}
where $\lambda$ is a balance parameter. The data term enforces alignment with the observation model, while the regularization term enforces the desired spectral image prior $R(\cdot)$.

To separate non-differentiable regularization terms from the data term in Eq.~\ref{solve_eq}, the variable splitting technique is often used, introducing an auxiliary variable $Z$ and reformulating the equation as a constrained optimization problem.
\begin{equation}
	\hat{I} = argmin{_I}{ \lvert\lvert{I_{rgb}-\mathcal{O}(I)}\rvert\rvert^2+\lambda R(Z)}, s.t. Z = I
\end{equation}

Afterward, the constrained optimization problem can be transformed into a nonconstrained optimization problem using the half quadratic splitting (HQS) method.
\begin{equation}
	(\hat{I}, \hat{Z})=\arg \min _{f, h}\|y-\mathcal{O} (I)\|^2+\eta\|Z-I\|^2+\lambda R(Z)
	\label{eq_noncon}
\end{equation}
where $\eta$ is a penalty parameter. Eq.~\ref{eq_noncon} can be split into two subproblems as
\begin{equation}
	I_k=argmin_I{\lvert\lvert{I_{rgb}-\mathcal{O}(I)}\rvert\rvert^2}+\eta\lvert\lvert{I-Z_{k-1}}\rvert\rvert^2 \label{i_k}
\end{equation}
\begin{equation}
	Z_k=argmin_Z{\mu \lvert\lvert{Z-I_k}\rvert\rvert^2+R(Z)} \label{z_k}
\end{equation}

The $I_k$-problem in Eq.~\ref{i_k} is a quadratic regularized least-squares problem that ensures the data fidelity. The direct solution can be given according to the specific encoding model.
\begin{equation}
	I_k = \arg \min _{I}\|I_{rgb}-\mathcal{O}(I)\|^2+\eta\left\|I-Z_{k-1}\right\|^2
	\label{model_driven}
\end{equation}

The regularization $Z_k$-problem in Eq.~\ref{z_k} is generally solved using a data-driven module with a deep neural network. This process can be expressed as:
\begin{equation}
	Z_k=f\left(I_k\right)
\end{equation}
Concretely, we adopt the Res-U-Net(depth=4) as the data-driven module.

\noindent {\bf{AEM.}} Rewrite Eq.\ref{model_driven} for AEM:
\begin{equation}
	I_k = \arg \min _{I}\|I_{rgb}-AID\|^2+\eta\left\|I-Z_{k-1}\right\|^2
\end{equation}
where A is the Mask of AEM and D represents the integration operation of the sensor. The solution is obtained using the gradient descent method:
\begin{equation}
	I_k = I_{k-1} - \alpha(-2AD^T(I_{rgb}-AI_{k-1})+2\eta(I_{k-1}-Z_{k-1}))
\end{equation}
where $\alpha$ is the step length of the gradient descent.

\noindent {\bf{PEM.}} Rewrite Eq.\ref{model_driven} for PEM: 
\begin{equation}
	I_k = \arg \min _{I}\|I_{rgb}-I\otimes P D \|^2+\eta\left\|I-Z_{k-1}\right\|^2
\end{equation}
where $P$ is the PSF of PEM and $D$ represents the integration operation of the sensor. 

\noindent The solution is obtained using the gradient descent method:
\begin{equation}
	I_k = I_{k-1} - \alpha(-2PD \otimes(I_{rgb}-I_{k-1} \otimes P D) + 2\eta(I_{k-1}-Z_{k-1}) )
\end{equation}
where $\alpha$ is the step length of the gradient descent.

\noindent {\bf{WEM.}} Rewrite Eq.\ref{model_driven} for WEM: 
\begin{equation}
	I_k = \arg \min _{I}\|I_{rgb}-I D \|^2+\eta\left\|I-Z_{k-1}\right\|^2
\end{equation}
where $D$ represents the integration operation of the sensor. The solution is obtained using the gradient descent method:
\begin{equation}
	I_k = I_{k-1}-\alpha(-2D^T(I_{rgb}-I_{k-1}D) + 2\eta(I_{k-1}-Z_{k-1}) )
\end{equation}
where $\alpha$ is the step length of the gradient descent.

\begin{table}[!t]
	\centering
	\caption{Comparison of different response curves of WEM-P, with the best results in bold.}
	\label{table_response}
	\resizebox{\linewidth}{!}
	{\begin{tabular}{lcccc}
			\hline
			\textbf{Response curves} & \textbf{PSNR$\uparrow$} & \textbf{PSNR-SI$\uparrow$} & \textbf{SAM$\downarrow$} & \textbf{ERGAS$\downarrow$} \\ \hline
			\textbf{FLIR\_BFS-U3-04S2C-C} & \textbf{44.96} & \textbf{38.81} & \textbf{0.03988} & \textbf{6.23} \\ 
			\textbf{Canon\_1DMarkIII} & 43.54 & 38.47 & 0.0461 & 7.88 \\ 
			\textbf{Canon\_20D} & 43.26 & 38.73 & 0.0472 & 8.38 \\ 
			\textbf{Canon\_300D} & 43.93 & 38.58 & 0.0438 & 7.23 \\ 
			\textbf{Canon\_40D} & 43.43 & 38.45 & 0.0481 & 8.09 \\ 
			\textbf{Canon\_500D} & 44.01 & 38.41 & 0.0443 & 7.25 \\ 
			\textbf{Canon\_50D} & 43.88 & 38.27 & 0.0445 & 7.34 \\ 
			\textbf{Canon\_5DMarkII} & 44.04 & 38.57 & 0.0435 & 7.37 \\ 
			\textbf{Canon\_600D} & 43.81 & 38.40 & 0.0452 & 7.66 \\ 
			\textbf{Canon\_60D} & 43.98 & 38.41 & 0.0437 & 7.35 \\ 
			\textbf{Hasselblad\_H2} & 44.18 & 39.00 & 0.0416 & 7.35 \\ 
			\textbf{Nikon\_D3X} & 43.85 & 38.75 & 0.0439 & 7.50 \\ 
			\textbf{Nikon\_D200} & 43.74 & 38.72 & 0.0444 & 7.64 \\ 
			\textbf{Nikon\_D3} & 43.47 & 38.55 & 0.0462 & 7.97 \\ 
			\textbf{Nikon\_D300s} & 43.65 & 38.68 & 0.0452 & 7.73 \\ 
			\textbf{Nikon\_D40} & 44.29 & 38.63 & 0.0426 & 6.69 \\ 
			\textbf{Nikon\_D50} & 44.31 & 38.76 & 0.0431 & 6.93 \\ 
			\textbf{Nikon\_D5100} & 43.33 & 38.57 & 0.0459 & 8.21 \\ 
			\textbf{Nikon\_D700} & 43.70 & 38.66 & 0.0444 & 7.90 \\ 
			\textbf{Nikon\_D80} & 43.24 & 38.55 & 0.0463 & 8.35 \\ 
			\textbf{Nikon\_D90} & 43.76 & 38.71 & 0.0441 & 7.63 \\ 
			\textbf{Nokia\_N900} & 44.05 & 38.52 & 0.0441 & 6.83 \\ 
			\textbf{Olympus\_E\_PL2} & 43.84 & 38.63 & 0.0436 & 7.37 \\ 
			\textbf{Pentax\_K\_5} & 43.93 & 38.63 & 0.0436 & 7.63 \\ 
			\textbf{Pentax\_Q} & 44.30 & 38.65 & 0.0437 & 6.97 \\ 
			\textbf{GS3-U3-50S5C-C} & 42.85 & 38.76 & 0.0542 & 9.30 \\ 
			\textbf{GS2-FW-14S5C-C} & 44.19 & 38.94 & 0.0425 & 7.35 \\ 
			\textbf{Phase\_One} & 43.41 & 38.09 & 0.0465 & 7.56 \\ 
			\textbf{SONY\_NEX\_5N} & 44.18 & 38.69 & 0.0439 & 7.05 \\ \hline
	\end{tabular}}
\end{table}

\section{Additional Experiments} \label{sec:ae}

We consider PSNR, PSNR-SI, SAM, and ERGAS for evaluation metrics. 
PSNR calculates the ratio of the maximum possible power of a signal to the power of corrupting noise that affects the fidelity.
PSNR-SI is the average PSNR of each channel of the spectral image, with intensity set to the channel maximum of the corresponding ground truth.
SAM calculates spectral similarity by measuring the angle between two spectral vectors.
ERGAS is the root mean square error ratio to average absolute spectral error.

\subsection{Response Selection of WEM-P} \label{sec:ad_response_selection}

We compare the response dataset~\cite{jiang2013space}, which includes 28 camera responses, to the response of the FLIR BFS\_U3\_04S2C\_C camera in the WEM-P. According to Table~\ref{table_response}, the latter experiment performs the best.
Therefore, we select the response curve of the FLIR BFS\_U3\_04S2C\_C camera as the response curve for the WEM-P and other experiments with fixed response curves.

\subsection{Initialization of Ideal Encoding Models}\label{sec:ad_init}
The ideal model encoding possesses significant flexibility, and the initialization considerably influences the experimental outcomes. Therefore, we conduct experiments on initializing AEM-I, PEM-I, and WEM-I encodings to determine the optimal initialization for the experimental setup. To this end, we employ two types of initialization: constant initialization and random initialization.

\begin{figure*}[!t]
	\centering
	\subfloat[Mask]{\includegraphics[width=4.1in]{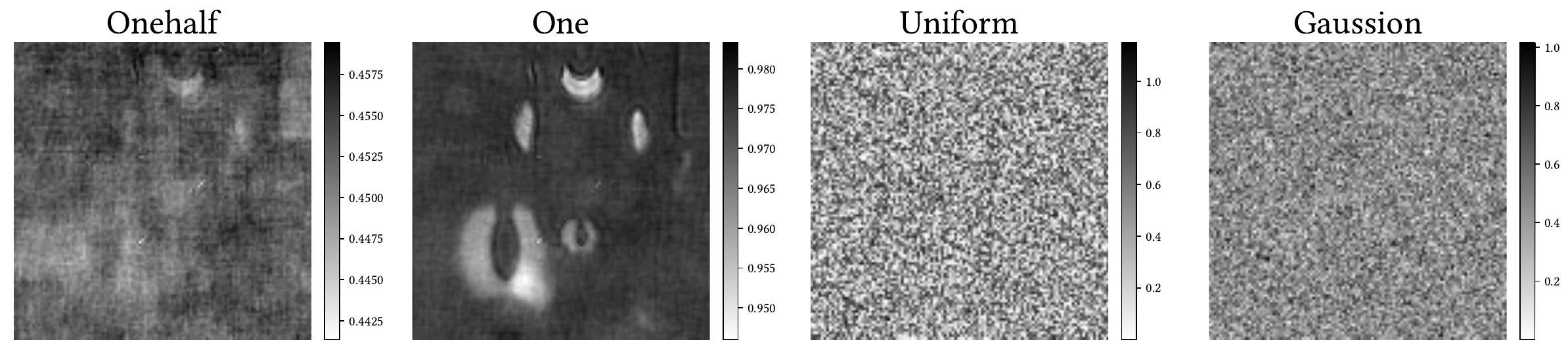} \label{AEMI_mask} }
	\subfloat[PSF]{\includegraphics[width=2.7in]{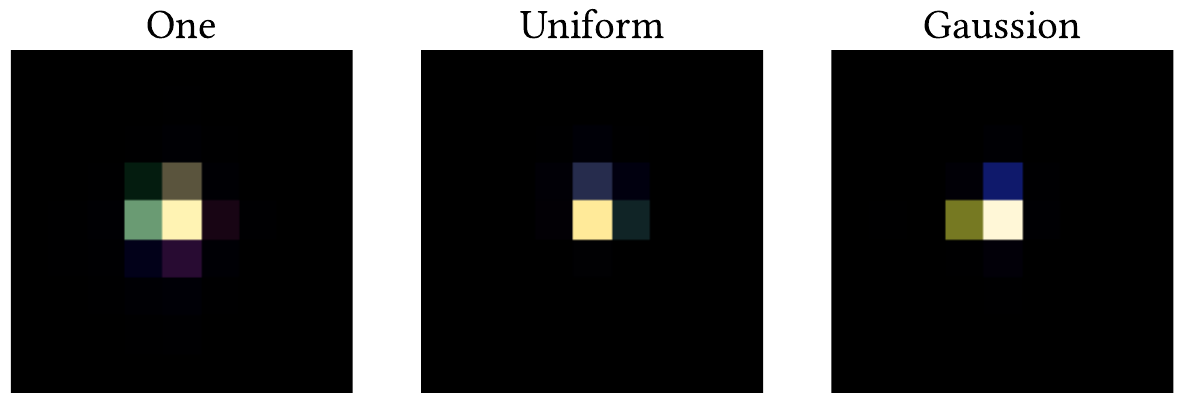}\label{pem_init}}
	
	\caption{Visual comparison of masks of AEM-I and PSFs of PEM-I. The mask is zoomed in for better visualization.}
	\label{fig_unit}
\end{figure*}
\begin{table}[!t]
	\centering
	\caption{Comparison of the different initializations of the AEM-I, with the best results in bold.}
	\label{table_a_initial}
	\resizebox{\linewidth}{!}
	{\begin{tabular}{ccccc}
			\hline
			\textbf{Initialization} & \textbf{PSNR$\uparrow$} & \textbf{PSNR-SI$\uparrow$} & \textbf{SAM$\downarrow$} & \textbf{ERGAS$\downarrow$} \\ \hline
			\textbf{One} & 44.64 & 38.58 & 0.0411 & 6.85 \\ 
			\textbf{Onehalf} & \textbf{45.08} & \textbf{38.95} & 0.0405 & 6.24  \\ 
			\textbf{Uniform} & 44.57 & 37.77 & \textbf{0.0394} & 5.84 \\ 
			\textbf{Gaussian} & 44.97 & 38.30 & 0.0407 & \textbf{5.79}
			\\ \hline
	\end{tabular}}
\end{table}

\noindent {\bf{Initialization of AEM-I.}} 
Two constant initialization types, all-0.5 and all-1, are used for AEM-I encoding. Two random initialization types are applied by setting the mask to a number between 0 and 1: uniform and Gaussian.

As shown in Table~\ref{table_a_initial}, the all-0.5 initialization setting has the highest PSNR and PSNR-SI test results, so we use the all-0.5 initialization as the experimental setup for the AEM-I. 
As no apparent features are observed after compositing the RGB image from the 31-channel mask, we select a channel mask for cropping to show the central region, as shown in Figure~\ref{AEMI_mask}. The constant-initialized mask is distinguishable from channel to channel, and the random-initialized mask remains morphologically randomly distributed.

\begin{table}[!t]
	\centering
	\caption{Comparison of the different initializations of the PEM-I, with the best results in bold.}
	\label{table_p_initial}
	\resizebox{\linewidth}{!}
	{\begin{tabular}{ccccc}
			\hline
			\textbf{Initialization} & \textbf{PSNR$\uparrow$} & \textbf{PSNR-SI$\uparrow$} & \textbf{SAM$\downarrow$} & \textbf{ERGAS$\downarrow$} \\ \hline
			\textbf{One} & 44.49 & 38.23 & 0.0419 & 6.67 \\ 
			\textbf{Uniform} & 44.87 & 38.78 & 0.0412 & 6.39 \\ 
			\textbf{Gaussian} & \textbf{44.90} & \textbf{38.68} & \textbf{0.0398} & \textbf{6.28} \\ \hline
	\end{tabular}}
\end{table}
\begin{table}[!t]
	\centering
	\caption{Comparison of the different initializations of the WEM-I w/ P.C., with the best results in bold.}
	\label{table_wem_init}
	\resizebox{\linewidth}{!}
	{\begin{tabular}{ccccc}
			\hline
			\textbf{Initialization} & \textbf{PSNR$\uparrow$} & \textbf{PSNR-SI$\uparrow$} & \textbf{SAM$\downarrow$} & \textbf{ERGAS$\downarrow$} \\ \hline
			\textbf{Random} & \textbf{44.26} & \textbf{38.84} & \textbf{0.0432} & \textbf{5.37} \\
			\textbf{Constant} & 42.19 & 35.62 & 0.0522 & 8.67 \\ \hline
	\end{tabular}}
\end{table}
\begin{figure}[!t]
	\centering
	{\includegraphics[width=3.2in]{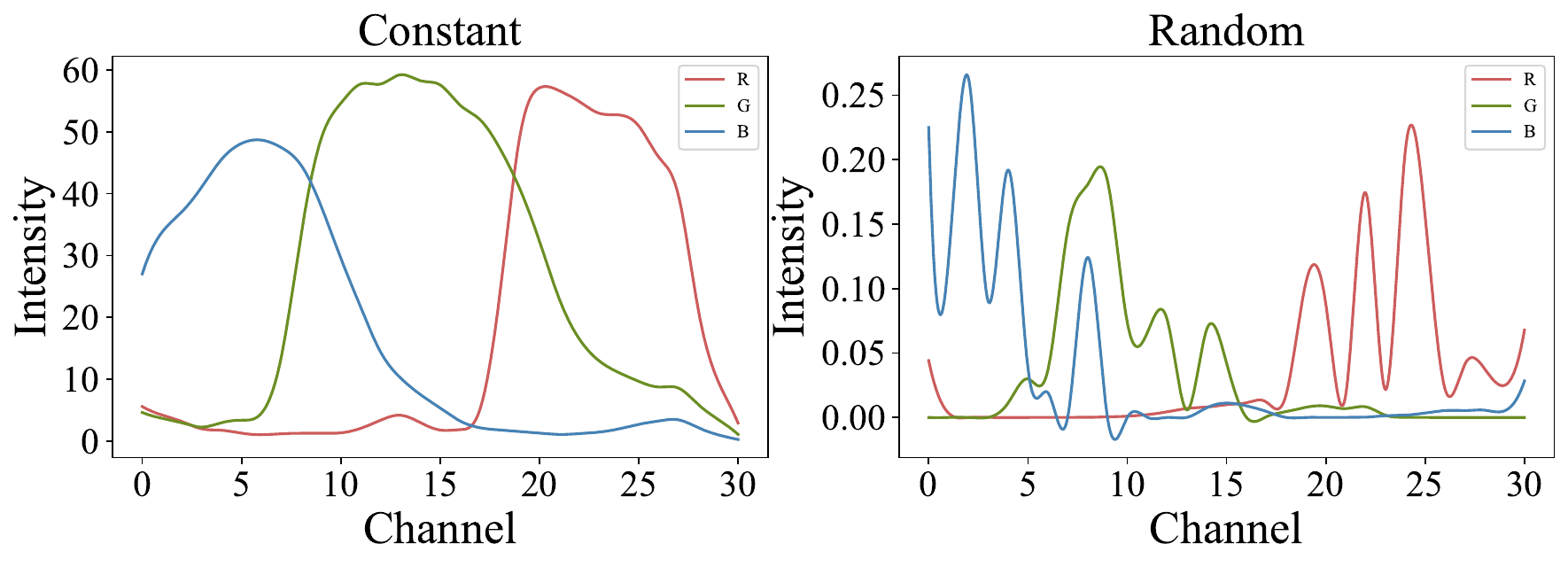}}
	\caption{Visual comparison of WEM-I w/ P.C. response curves with different initializations.}
	\label{fig_wem_init}
\end{figure}

\noindent {\bf{Initialization of PEM-I.}}
We utilize a constant initialization of all ones and uniform and Gaussian random initialization for PEM-I.

Table~\ref{table_p_initial} concludes that the PEM-I model performs best with Gaussian random initialization for all evaluation metrics. Therefore, we select the Gaussian random initialization as the experimental setup for the PEM-I. Figure~\ref{pem_init} shows a composite RGB image of the 31-channel PSFs obtained using this initialization, which exhibits a central convergence pattern that leads to the best imaging quality.

\noindent {\bf{Initialization of WEM-I.}} 
In WEM-I, a linear layer is employed for encoding instead of a fixed RGB filter response function. 
We try constant initialization of the response curve of the FLIR BFS\_U3\_04S2C\_C camera and random initialization to initialize the weights of the convolution kernel in the WEM-I with positive constraints  (\mbox{WEM-I w/ P.C.}).

Table~\ref{table_wem_init} illustrates that random initialization yields better imaging performance than constant initialization. Moreover, the response curve for constant initialization remains unchanged from the initialization settings, while the response curve for random initialization has a distinctive shape, as shown in Figure~\ref{fig_wem_init}. Consequently, random initialization is used in WEM-I and WEM-I w/ P.C experiments.

\subsection{Selection of Loss Functions}\label{sec:ae_loss}
In addition to the MAE loss function used in the main text experiments, we also compare the effects of two additional loss functions on the results: MSE loss and ERGAS loss.

\begin{table*}[!t]
	\centering
	\caption{Spectral imaging performance of systems using UEMs on different loss functions, with the best results in bold.}
	\label{table_loss}
	\resizebox{\linewidth}{!}{
		\begin{tabular}{c|c|cccc|c|cccc}
			\hline
			~ & \multicolumn{5}{c|}{\textbf{Physical Model}}& \multicolumn{5}{c}{\textbf{Ideal Model}} \\ \hline
			\textbf{Loss} & \textbf{Encoding Model} & \textbf{PSNR$\uparrow$} & \textbf{PSNR-SI$\uparrow$} & \textbf{SAM$\downarrow$} & \textbf{ERGAS$\downarrow$} & \textbf{Encoding Model} & \textbf{PSNR$\uparrow$} & \textbf{PSNR-SI$\uparrow$} & \textbf{SAM$\downarrow$} & \textbf{ERGAS$\downarrow$}  \\ \hline
			\multirow{3}*{\textbf{MAE}} & \textbf{WEM-P} & 44.96 & 38.81 & 0.0399 & 6.23 & \textbf{WEM-I} & \textbf{45.25} & \textbf{38.51} & \textbf{0.0328} & \textbf{5.23}  \\ 
			~ & \textbf{AEM-P} & 40.21 & 32.66 & 0.0491 & 8.96 & \textbf{AEM-I} & 44.64 & 38.58 & 0.0411 & 6.85  \\ 
			~ & \textbf{PEM-P} & 33.75 & 26.02 & 0.0655 & 16.62 & \textbf{PEM-I} & 44.9 & 38.68 & 0.0398 & 6.28  \\  \hline
			\multirow{3}*{\textbf{MSE}} & \textbf{WEM-P} & \textbf{43.94} & \textbf{37.74} & 0.0465 & 6.73 & \textbf{WEM-I} & 43.52 & 36.77 & \textbf{0.0448} & \textbf{6.63}  \\ 
			~ & \textbf{AEM-P} & 38.41 & 30.78 & 0.0611 & 10.47 & \textbf{AEM-I} & 43.48 & 37.2 & 0.0485 & 7.03  \\ 
			~ & \textbf{PEM-P} & 33.71 & 25.99 & 0.0689 & 16.7 & \textbf{PEM-I} & 43.16 & 36.75 & 0.0498 & 7.15  \\  \hline
			\multirow{3}*{\textbf{ERGAS}} & \textbf{WEM-P} & \textbf{44.49} & \textbf{38.24} & \textbf{0.0427} & 6.3 & \textbf{WEM-I} & 44.11 & 38.29 & 0.0428 & \textbf{5.14}  \\ 
			~ & \textbf{AEM-P} & 39.9 & 32.37 & 0.0523 & 8.98 & \textbf{AEM-I} & 43.77 & 38.12 & 0.0448 & 5.4  \\ 
			~ & \textbf{PEM-P} & 34.14 & 26.77 & 0.0690 & 15.99 & \textbf{PEM-I} & 44.43 & 38.27 & 0.0430 & 6.43  \\ \hline
	\end{tabular}}
\end{table*}

Table~\ref{table_loss} demonstrates that switching the loss function has no impact on the consistency of individual systems. MAE loss is the most appropriate loss function for training joint encoder-decoder optimization computational spectral imaging systems, compared to MSE loss and ERGAS loss. Therefore, we select the MAE loss as the loss function for the experiments in the main text. When MSE loss is used as the loss function, the PEM-I encounters convergence problems. This is because MSE loss is not well-suited for models with a high degree of freedom. ERGAS loss provides only a slight improvement in the ERGAS metric.

\subsection{PSF Sizes and Kernel Sizes of PEM-I}\label{sec:ae_kernel}

\begin{figure}[!t]
	\centering
	\captionsetup[subfloat]{labelsep=none,format=plain,labelformat=empty}
	\subfloat[\scriptsize{Kernel Size: 3}]{\begin{overpic}[width=0.7in]{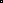}
			\put(-20,10){\scriptsize\rotatebox{90}{\color{black} PSF Size: 3}}
	\end{overpic}}
	\label{conv_patch_res_sim_3k_3p}
	\subfloat[\scriptsize{Kernel Size: 5}]{\includegraphics[width=0.7in]{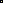}}
	\label{conv_patch_res_sim_5k_3p}
	\subfloat[\scriptsize{Kernel Size: 7}]{\includegraphics[width=0.7in]{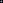}}
	\label{conv_patch_res_sim_7k_3p}
	\subfloat[\scriptsize{Kernel Size: 9}]{\includegraphics[width=0.7in]{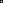}}
	\label{conv_patch_res_sim_9k_3p}
	
	\subfloat[\scriptsize{Kernel Size: 3}]{\begin{overpic}[width=0.7in]{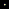}
			\put(-20,10){\scriptsize\rotatebox{90}{\color{black} PSF Size : 9}}
	\end{overpic}}
	\label{conv_patch_res_sim_3k_9p}
	\subfloat[\scriptsize{Kernel Size: 5}]{\includegraphics[width=0.7in]{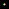}}
	\label{conv_patch_res_sim_5k_9p}
	\subfloat[\scriptsize{Kernel Size: 9}]{\includegraphics[width=0.7in]{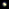}}
	\label{conv_patch_res_sim_9k_9p}
	\subfloat[\scriptsize{Kernel Size:15}]{\includegraphics[width=0.7in]{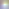}}
	\label{conv_patch_res_sim_15k_9p}
	
	\caption{Visual comparison of RGB PSFs of PEM-I with different sizes of learnable PSF and different convolutional kernel sizes of the decoding model.}
	\label{fig_FreePSF_size}
\end{figure}

We conduct experiments with varying PSF sizes and convolutional kernel sizes of the decoding model. The comparison results are shown in Table~\ref{table_p_psfkernel} and Figure~\ref{fig_FreePSF_size}. The results indicate that smaller learnable PSF and kernel sizes in the decoding model can lead to greater convergence of PSF and improve performance. 


In summary, to achieve clear imaging quality in the PEM, a concentrated PSF must be designed while maintaining a slight variance of the PSF across the channels to preserve spectral information.

\begin{table}[!t]
	\centering
	\caption{Comparison of the imaging quality of systems using the PEM-I and the Sim-Conv-Net with different PSF and kernel size. The best result is marked in bold, and the second-best result is underlined in each column.}
	\label{table_p_psfkernel}
	\resizebox{\linewidth}{!}{
		\begin{tabular}{ccccc}
			\hline
			\textbf{PSF Size} & \textbf{Kernel Size} & \textbf{PSNR$\uparrow$} & \textbf{PSNR-SI$\uparrow$} & \textbf{SAM$\downarrow$} \\ \hline
			3 & 3 & \textbf{44.75} & \textbf{38.69} & \textbf{0.0409} \\ 
			3 & 5 & 44 & 37.54 & 0.0421 \\ 
			3 & 7 & 42.43 & 35.57 & 0.0471 \\ 
			3 & 9 & 41.54 & 34.35 & 0.0483 \\ 
			9 & 3 & \underline{44.55} & \underline{38.32} & \underline{0.0417} \\ 
			9 & 5 & 44.01 & 37.47 & 0.0421 \\ 
			9 & 9 & 40.28 & 32.91 & 0.0512 \\ 
			9 & 15 & 34.14 & 26.52 & 0.0776 \\ 
			16 & 3 & 44.4 & 38.05 & 0.0419 \\ 
			16 & 5 & 40.43 & 33.33 & 0.0568 \\ 
			64 & 3 & 40.6 & 33.24 & 0.0509 \\ \hline
	\end{tabular}}
\end{table}

\subsection{Full Optimization}\label{sec:ae_uem}
We combine UEM-I with a neural network as a decoding model to construct a fully optimized ideal computational spectral imaging system. This system can simultaneously optimize mask, PSF, spectral response, and neural networks.

As shown in Table~\ref{table_withuem}, UEM-I outperforms WEM-I in PSNR metrics only, and there is no significant improvement in imaging performance compared to single-mode free encoding. Therefore, we suggest that restricting optimization to a single free encoding is adequate for approaching the upper-performance limit of all free encoding.

\begin{table}[!t]
	\centering
	\caption{Comparison of the imaging quality of systems using different ideal models of UEM and Sim-Conv-Net, with the best results in bold and the next best results underlined.}
	\label{table_withuem}
	\resizebox{\linewidth}{!}
	{\begin{tabular}{ccccc}
			\hline
			\textbf{Encoding Model} & \textbf{PSNR$\uparrow$} & \textbf{PSNR-SI$\uparrow$} & \textbf{SAM$\downarrow$} & \textbf{ERGAS$\downarrow$} \\ \hline
			\textbf{UEM-I} & \textbf{45.30} & 38.48 & \underline{0.0372} & \underline{5.63} \\ 
			\textbf{WEM-I} & \underline{45.25} & 38.51 & \textbf{0.0328} & \textbf{5.23} \\ 
			\textbf{AEM-I} & 45.08 & \textbf{38.95} & 0.0405 & 6.24 \\ 
			\textbf{PEM-I} & 44.9 & \underline{38.68} & 0.0398 & 6.28 \\ \hline
	\end{tabular}}
\end{table}

\begin{table}[!t]
	\centering
	\caption{Spectral imaging performance of systems using the WEM-P on ICVL datasets.}
	\label{table_data}
	\resizebox{\linewidth}{!}{
		\begin{tabular}{ccccccc}
			\hline
			\textbf{Dataset} & \textbf{PSNR$\uparrow$} & \textbf{PSNR-SI$\uparrow$} & \textbf{SAM$\downarrow$} & \textbf{ERGAS$\downarrow$} \\ \hline
			ICVL & 44.96 & 38.81 & 0.0399 & 6.23 \\ 
			ICVL-90\%Training set & 44.77 & 38.76 & 0.0402 & 6.36 \\ 
			ICVL-80\%Training set & 44.18 & 38.36 & 0.0437 & 6.59 \\ 
			ICVL-Redividing & 44.65 & 38.65 & 0.0413 & 6.49 \\ \hline
	\end{tabular}}
\end{table}

\subsection{Validation of ICVL Dataset}\label{sec:ae_icvl}
During training, we use the common practice of cropping each image into multiple 512 × 512 patches, with sufficient data to cover the training requirements. We conduct three additional experiments under the WEM-P setting, reducing the training set by 10\% and 20\%, and repartitioning the training and validation sets. As shown in Table~\ref{table_data}, the fluctuation of the test results is within 1 dB for both reducing the amount of data in the training set and repartitioning the dataset, proving the validity of the training data.

\section{Visualization of Ideal Encoding Patterns}\label{sec:v}
\begin{figure*}[p]
	\centering
	\includegraphics[width=\textwidth,height=\textheight,keepaspectratio]{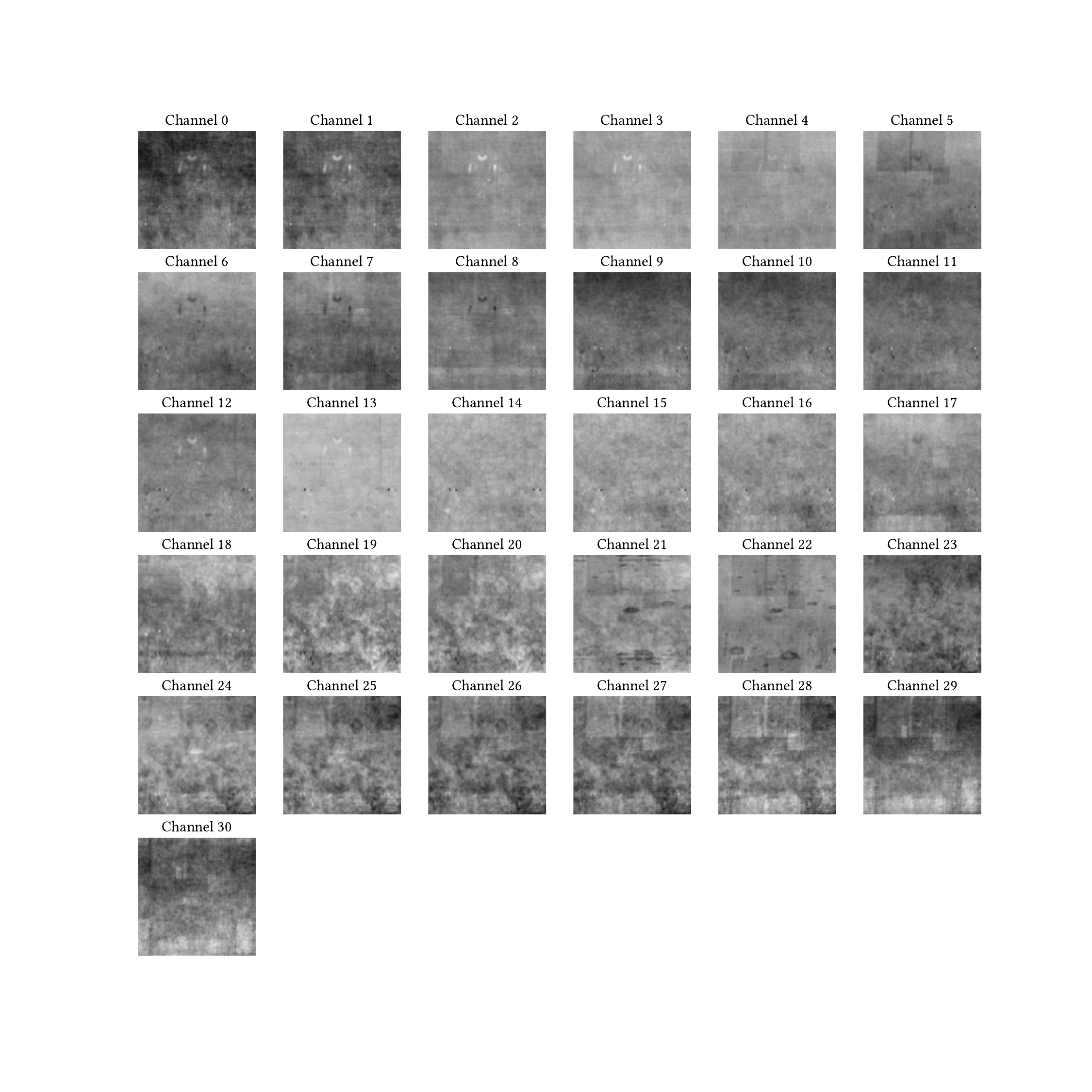}
	\caption{Visualization of the AEM-I masks.}
	\label{fig_AEMI_mask}
\end{figure*}

\begin{figure*}[p]
	\centering
	\includegraphics[width=\textwidth,height=\textheight,keepaspectratio]{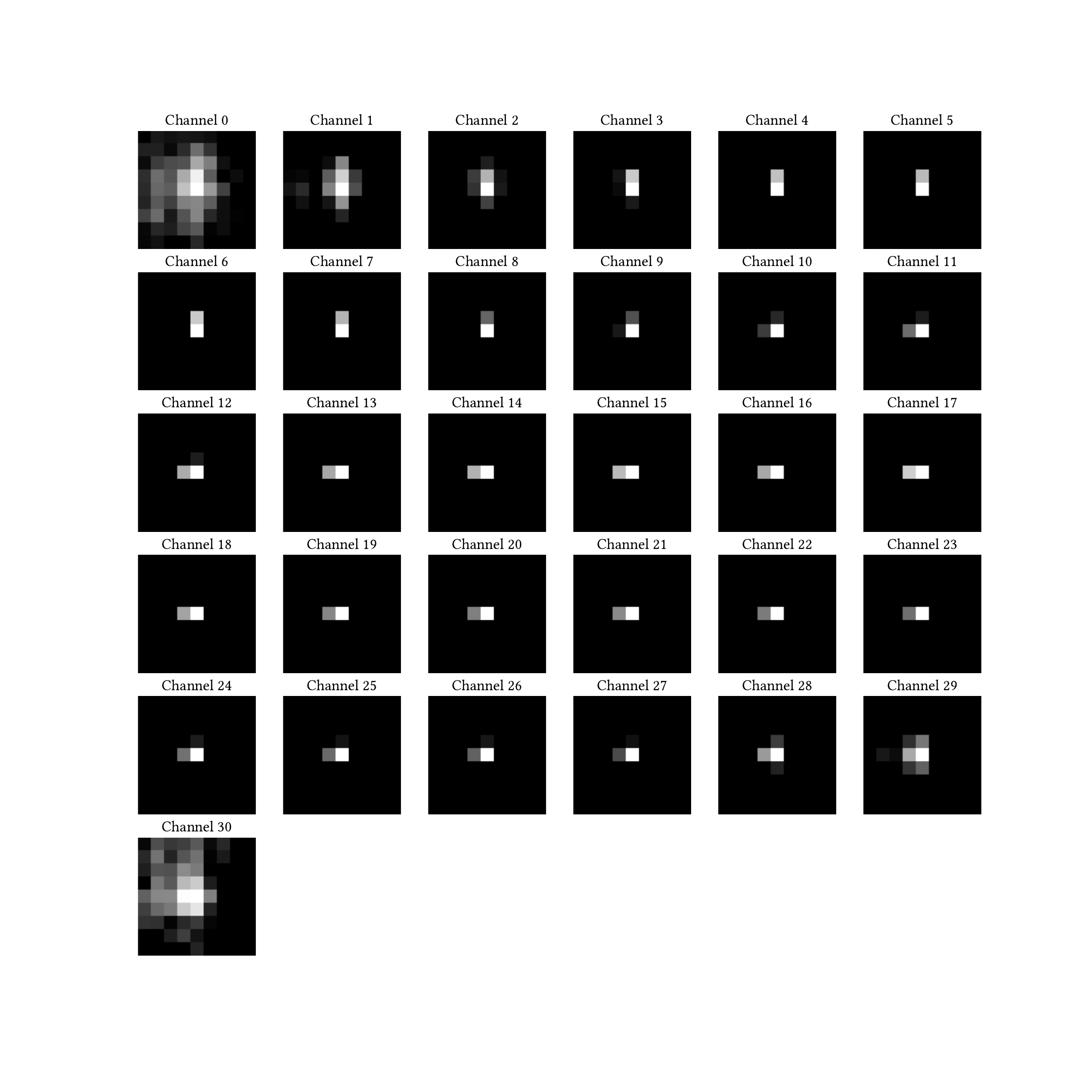}
	\caption{Visualization of the PEM-I PSFs.}
	\label{fig_PEMI_psf}
\end{figure*}

%
Due to the limited space, we demonstrate a few channels of PSF and mask in the main paper. We further provide the full version here. Figure~\ref{fig_AEMI_mask} shows the full visualization of the 31 channels of the AEM-I masks. Figure~\ref{fig_PEMI_psf} shows the full visualization of the 31 channels of the PEM-I PSFs.

{
    \small
    \bibliographystyle{ieeenat_fullname}
    \bibliography{main}
}
